\documentclass[11pt]{article}
\pdfoutput=1 
\usepackage{jheppub} 

\usepackage{amsmath,amssymb,amsthm,amscd}

\usepackage[dvipsnames]{xcolor}

\definecolor{darkblue}{rgb}{0.1,0.1,.7}
\usepackage{graphicx}

\usepackage{tikz-cd}

\usepackage{dsdshorthand}
\usepackage{mathtools} 
\usepackage{bbm}

\numberwithin{equation}{section}

\newcommand{\tr}{{\rm Tr}}
\newcommand{\beq}{\begin{equation}}
\newcommand{\eeq}{\end{equation}}

\newcommand{\trans}{{\sf T}}

\renewcommand{\be}{\begin{eqnarray}}
\renewcommand{\ee}{\end{eqnarray}}
\newcommand{\bea}{\begin{eqnarray}}
\newcommand{\eea}{\end{eqnarray}}

\newcommand{\nn}{\nonumber}

\newcommand{\PTerm}[2]{\big\{{\textstyle\genfrac{}{}{0pt}{}{#1}{#2}}\big\}}

\newcommand{\unit}{\mathbbm{1}}

\def\XXint#1#2#3{{\setbox0=\hbox{$#1{#2#3}{\int}$}
     \vcenter{\hbox{$#2#3$}}\kern-.5\wd0}}

\newcommand{\remarkjc}[1]{{\renewcommand{\bfdefault}{b}{\color[RGB]{0,0,150}{\textbf{#1}}}}}
\providecommand{\remarkjc}[1]{\ignorespaces}

\title{Maximally Supersymmetric RG Flows in 4D and Integrability}

\author[\hspace{-.225em}a,b]{Jo\~ao~Caetano,}
\author[\hspace{-.225em}c]{Wolfger Peelaers,}
\author[\hspace{.08em}a]{Leonardo Rastelli}

\affiliation[a]{C. N. Yang Institute for Theoretical Physics, Stony Brook University, Stony Brook, NY 11794, USA}
\affiliation[b]{Simons Center for Geometry and Physics, Stony Brook University, Stony Brook, NY 11794, USA }
\affiliation[c]{Mathematical Institute, University of Oxford, Woodstock Road, Oxford, OX2 6GG, United Kingdom}

\preprint{YITP-SB-20-12}

\abstract{
We revisit the leading irrelevant deformation of $\mathcal{N}=4$ Super Yang-Mills theory that preserves sixteen supercharges. We consider the deformed theory on $S^3 \times \mathbb{R}$.
We are able to write  a closed form expression of the classical action  thanks to a formalism that realizes eight supercharges off shell.
We then investigate
  integrability of the spectral problem,
   by studying the spin-chain Hamiltonian in planar perturbation theory. 
 While there are some structural indications that a suitably defined deformation might preserve integrability, we are unable to settle this question by  our 
    two-loop calculation; indeed up to this order we recover the integrable Hamiltonian of undeformed $\mathcal{N}=4$ SYM due to accidental symmetry enhancement.
    We also  comment on the holographic interpretation of the  theory.}

\begin{document}
\maketitle

\flushbottom


\newpage

\section{Introduction}

In this paper we revisit the 
 deformation of ${\cal N}=4$ super Yang-Mills theory 
 by  the leading {\it irrelevant} operator  that preserves
16 of the total 32 supercharges.  
 Any maximally supersymmetric RG flow whose endpoint is ${\cal N}=4$ SYM can be described in the infrared
  by an effective action of the form
  \begin{equation} \label{EFT8}
S_{ \rm SYM} + h \int d^4x \,  {\cal O}_8 + \dots  \, ,
\end{equation}
  where ${\cal O}_8$ denotes the leading irrelevant scalar operator (of conformal dimension $\Delta = 8$, as it turns out) 
  and  the dots indicate an infinite tower of higher-dimensional operators. 
  
\medskip

\noindent 
{\it Examples and inspiration}

\noindent
  There are several interesting examples of RG flows that end in (\ref{EFT8}).  A familiar one is the  Coulomb branch flow, obtained by moving on the moduli space of vacua
of an $SU(N+k)$  SYM gauge theory in such a way that an $SU(N) \times U(1)^k$ subgroup remains unbroken. In the far infrared, the theory is described by $k$ decoupled vector multiplets, plus an interacting superconformal field theory (SCFT),
with a leading irrelevant correction given by  ${\cal O}_8$. In this case, the UV fixed point is a well-defined four-dimensional field theory, indeed another SYM theory with larger gauge group.
Another class of examples arise by considering flows from higher-dimensional SCFTs, such as  compactification of  $(2, 0)$ theories on the two-torus, yielding ${\cal N}=4$ SYM at low energy. 
Perhaps the most intriguing maximally supersymmetric flow is the one defined by the full open string field theory on $N$ D3 branes in IIB string theory. On general grounds, the infrared theory takes again the form (\ref{EFT8}), with $h \sim (\alpha')^2$, but the UV behavior is not controlled by a  local field theory. 

It has been a long-standing speculation that  the canonical AdS/CFT duality (the  equivalence of ${\cal N} = 4$ SYM theory and $AdS_5 \times S^5$ string theory) might extend  beyond the low energy/near horizon limit -- that the full D3 brane effective field theory might be dual to closed string theory in the full asymptotically flat D3 brane geometry. In fact, this idea even predates  the precise formulation~\cite{Maldacena:1997re, Gubser:1998bc, Witten:1998qj} of  AdS/CFT, being implicit in the comparisons of worldvolume and gravity calculations of 
particle absorption by D-branes by Klebanov and collaborators \cite{Klebanov:1997kc, Gubser:1997yh, Gubser:1997se} that provided  crucial early hints for the gauge/gravity correspondence.  A more explicit statement of this conjecture was given by Intriligator \cite{Intriligator:1999ai}.  Consider closed string theory on the background  defined by the three-brane metric 
\begin{eqnarray} \label{D3metric}
ds^2 & = & H^{-\frac{1}{2}} dx_m dx_m  + H^{\frac{1}{2}}  dx_I dx_I  \quad \\
H (r) & = & \tilde h + \frac{R^4}{r^4} \,,   \quad r^2 = x_I x_I \, , \nonumber
\end{eqnarray}
with coordinates $x_m$ spanning the brane worldvolume $\mathbb{R}^{3, 1}$ and coordinates $x_I$ spanning the transverse $\mathbb{R}^6$. The scale $R$ is fixed by $R^4 = 4 \pi g_s N (\alpha')^4$,
where $N$ is the number of D3 branes, while $\tilde h$ is an arbitrary integration constant. There are also $N$ units of self-dual five form flux $F_5 \sim (1 + *)  (dx)^4 \wedge dH^{-1}$.
Intriligator argued that  this closed string  background
is dual to  the boundary theory defined by the action
\begin{equation} \label{Ken}
S_{ \rm SYM} + \tilde h R^4    \int d^4x \, {\cal O}_8  \,. 
\end{equation}
In order to test this intriguing proposal, we need a more precise understanding of the field theory side. Is (\ref{Ken}) meant as a shorthand for the full open string field theory? Or is there a sense in which (due perhaps to the high degree of supersymmetry, or some other principle) one can make sense of it in field-theoretic terms,  unambiguously moving upstream the RG flow? These difficult questions are immaterial to leading order
in $\tilde h$, and remarkably, the proposal passes a non-trivial leading-order test in the supergravity limit of large $N$ and large 't Hooft coupling \cite{Rastelli:2000xj}.

We feel  encouraged to revisit this circle of ideas thanks to the recent  progress in understanding a very special irrelevant deformation of 
  two-dimensional QFTs, the so-called  $T \bar T$ deformation \cite{Zamolodchikov:2004ce,Smirnov:2016lqw, Cavaglia:2016oda}  (see also \cite{Jiang:2019hxb} for a review). 
  Due to its ``quasi-topological'' nature, the $T \bar T$ deformation appears to define a sensible theory, albeit one with  exotic UV behavior,
   outside the framework of conventional local quantum field theory. This is just what one would desire for the flow~(\ref{Ken}).  Is there perhaps a sense in which ${\cal O}_8$ may be regarded
   as a four-dimensional, maximally supersymmetric analog of  $T \bar T$?

 \medskip 
  
\noindent 
{\it The leading irrelevant deformation}

\noindent
  Let us now describe the leading irrelevant operator  ${\cal O}_8$ in some more detail, and review some of  the requisite  superconformal representation theory.
 The systematic classification of deformations of ${\cal N}=4$ SCFTs that preserve maximal supersymmetry has been carried out   in \cite{Cordova:2016xhm}, as a special case of a much more general analysis. 
 The main result is easy to understand in elementary terms \cite{Intriligator:1999ai}.  To preserve maximal supersymmetry, we need to add the top component of an ${\cal N}=4$ multiplet. As we are after the susy-preserving deformations of lowest dimension,
 we should focus on the shortest (one-half BPS) multiplets. The one-half BPS multiplets are built on superconformal primary operators ${\cal O}^{(p)}$, scalars  of dimension $\Delta = p$, $p \geq 2$,
 in the $p$-th symmetric traceless representation of the $SO(6)$ R-symmetry. The susy-preserving deformation of smallest dimension is  the top components of the $p=2$ multiplet (the stress-tensor supermultiplet),  ${\cal O}_\tau = Q^4 {\cal O}^{(2)}$. This is 
  a complex scalar operator  with $\Delta =4$, corresponding to the
 exactly marginal deformation parametrized by the complexified gauge coupling $\tau$. The $p=3$ multiplet  
 does not contain a scalar among its top components. 
Finally, the top component of the  $p=4$ multiplet 
is a real scalar, R-symmetry singlet of dimension eight,  ${\cal O}_8 =  Q^4 \tilde Q^4  {\cal O}^{(4)}$,
 which is thus the leading irrelevant deformation of an ${\cal N}=4$ SCFT.  

The abstract representation-theoretic argument that we have just reviewed is blind to the color structure of the theory.
In ${\cal N}=4$ SYM  with gauge group $SU(N)$ (for $N>2$) there are two linearly independent versions of  ${\cal O}_8$, a single-trace and a double-trace version. The single-trace version reads
\bea
{\cal O}^{\rm ST}_8  & = & Q^4 \tilde Q^4 \, {\rm Tr}\,  \Phi^{(I}  \Phi^J  \Phi^K  \Phi^{L)} \\ 
& =& {\rm Tr} \Big[F^4 - \frac{1}{4} (F^2)^2+  4\big(F_{mp}F^{np} - \tfrac{1}{4} F_{pq}F^{pq}\delta_{m}^{n}\big) D^m \Phi_I D_n \Phi_I \nonumber \\
&&\qquad -(D_m\Phi_I)(D^m\Phi_I)(D_n\Phi_J)(D^n\Phi_J) +2 (D_m\Phi_I)(D^m\Phi_J)(D_n\Phi_I)(D^n\Phi_J) + ... \Big] \nonumber
\eea
where $\Phi^I$, $I = 1, \dots 6$ are the six real scalar fields and $F_{mn}$ the gauge field strength. Terms containing fermions are omitted. We recognize the structure of the leading irrelevant operator in the derivative expansion of the Dirac-Born-Infeld action for D3 branes (see \cite{Tseytlin:1999dj} for a nice review). This has to be, since the  D3 brane flow is maximally supersymmetric, and as we have just argued ${\cal O}^{\rm ST}_8$ is the unique leading irrelevant single-trace deformation that preserves the full Poincar\'e supersymmetry. 

On the other hand, the double-trace version takes the schematic form
\be
{\cal O}^{\rm DT}_8 = Q^4 \tilde Q^4  \, {\rm Tr}\, \Phi^{(I}  \Phi^J \, \tr\, \Phi^K  \Phi^{L)} = T_{mn} T_{mn}  +  {\cal O}_\tau  {\cal O}_{\bar \tau} + \dots
\ee
Superficially, this comes as close to a four-dimensional maximally supersymmetric version of the two-dimensional
 $T \bar T$ operator  as one may hope, but there is no reason to expect any 
 of the same remarkable quasi-topological properties.\footnote{For very different approaches to  higher-dimensional generalizations of $T \bar T$, see, e.g., \cite{Taylor:2018xcy,Hartman:2018tkw,Belin:2020oib}.}

The complete list of irrelevant deformations of ${\cal N}=4$ SCFTs that preserve maximal supersymmetry can be found  in Table 28 of \cite{Cordova:2016xhm}. There are four classes: three based on shortened representations
 (one-half BPS, one-quarter BPS and one-sixteen BPS) and one based on  long representations. If one further demands that the deformation be an R-symmetry singlet, so that the full $SU(4)_R$ is preserved,
 the list is shorter: besides ${\cal O}_8$, there is one deformation of dimension 10, top component of a one-quarter BPS multiplet,
 and an infinite set of non-protected deformations with $\Delta > 10$. The one-quarter BPS deformation of dimension 10 is realized in  ${\cal N}=4$ SYM by a double-trace operator. 
  If we  are interested in the planar limit of the theory, we should restrict attention 
 to single-trace deformations, and then
  ${\cal O}^{\rm ST}_8$ is the {\it unique} protected irrelevant deformation that preserves the full Poincar\'e supersymmetry and the full R-symmetry.  
  
  These considerations provide a first way to interpret Intriligator's proposal.
   In the limit of large 't Hooft coupling $\lambda = g_{\rm YM}^2 N$, 
   where ${\cal N}=4$ SYM is dual to the classical supergravity background (\ref{D3metric}),
   all long single-trace operators  (dual to single string states) acquire a large anomalous dimension. There is then at least a formal sense that
  we are perturbing   the theory by just   ${\cal O}^{\rm ST}_8$, as indicated in (\ref{Ken}). At finite $\lambda$ however, representation theory  alone does not determine the dots in (\ref{EFT8}), and we need a principle to fix the infinite tower of D-terms  $Q^8 \tilde Q^8  {\cal O}^{\rm L}_{\Delta}$, where   ${\cal O}^{\rm L}_{\Delta}$ is a Lorentz scalar, R-symmetry singlet superprimary, with $\Delta > 2$ from 
  unitarity.\footnote{The lightest such operator is the Konishi operator ${\rm Tr}\, \Phi_I \Phi_I$.} The same ambiguity plagues all naive attempts to fix the non-abelian version of the Dirac-Born-Infeld action for D3 branes.

\medskip

 \noindent 
{\it Integrability on $S^3 \times \mathbb{R}$?}
 \nopagebreak[4]
 
\noindent
In this paper, we will focus on the planar limit of ${\cal N}=4$ SYM theory and thus restrict attention to  single-trace irrelevant deformations. This is similar in spirit to \cite{Giveon:2017nie, Giveon:2017myj, Asrat:2017tzd},
 which studied  a single-trace version  of the $T \bar T$ deformation in  two-dimensional conformal field theory examples.\footnote{The analogy between  ${\cal O}^{\rm ST}_8$ and the deformation
 considered in  \cite{Giveon:2017nie, Giveon:2017myj, Asrat:2017tzd} becomes sharper if one looks at  their $(4, 4)$ supersymmetric examples.
 Superconformal representation theory  works  rather similarly in the $2d$  $(4, 4)$ case and in the 
 $4d$ ${\cal N}=4$ case. In both cases, the leading irrelevant deformation is the top component of a one-half BPS multiplet, whose superprimary has R quantum numbers
{\it twice} those of the stress-tensor multiplet. Supersymmetric extensions of  the $T \bar T$ deformation have been studied in, e.g., \cite{Baggio:2018rpv,Chang:2018dge,Jiang:2019hux,Chang:2019kiu,Coleman:2019dvf,Brennan:2019azg}.} 

The $T \bar T$ deformation  \cite{Smirnov:2016lqw,Cavaglia:2016oda} 
can be canonically defined all along the RG flow, for  finite values of the deformation parameter $h$. A key step is the derivation of  differential equations for physical observables as functions of $h$, which can then be integrated starting from the IR initial conditions. 
We do not have such a luxury, but we can still attempt to make sense of our irrelevant deformation order by order in conformal perturbation theory, as a series expansion in $h$. Of course, we need to cancel
divergences by adding suitable local counterterms, which  as argued above take the form of D-terms  $Q^8 \tilde Q^8  {\cal O}^{\rm L}_{\Delta}$.
Barring some additional principle, the finite part of these counterterms is ambiguous, with the number of undetermined coefficients growing at each successive order in $h$.

Could integrability be the missing principle? Famously,  planar ${\cal N}=4$ SYM theory is integrable \cite{Beisert:2010jr}. Integrability is so powerful to lead, at least in principle, to a complete solution of the theory \cite{Gromov:2009zb,Bombardelli:2009ns,Arutyunov:2009ur}. We are speculating that among the infinitely many RG trajectories
that look in the IR as ${\cal N}=4$ SYM perturbed by ${\cal O}^{\rm ST}_8$, there may be one (and perhaps many) that preserve planar integrability. 
The question of  planar integrability of irrelevant deformations of ${\cal N}=4$ SYM was initially motivated by the desire to make sense of the dual field theory to full D3 brane geometry,  but is worth investigating in its own right. The present  paper is a first exploration of this question. 

Our approach will be to study the theory on the cylinder ${S^3} \times \mathbb{R}$, because this setup allows for a particularly sharp formulation
of the integrability question.
At the conformal point  the theory on the cylinder is entirely equivalent to the theory on $\mathbb{R}^4$, because the two frames are related
 by a Weyl transformation. As we turn on the irrelevant deformation, the two frames cease to be equivalent. We will be concerned with the spectral problem on the cylinder,
 i.e., with the evolution of energy eigenstates as a function of the dimensionless deformation parameter $h/ \ell^4$, where $\ell$ is the radius of the  ${S^3}$.  For  $h =0$,
 we have the usual state/operator map of conformal field theory: states on $S^3$ are in one-to-one correspondence with local operators, with the cylinder Hamiltonian being identified with the flat-space dilation generator.
As  we turn on the deformation, the state/operator map is lost, but we still have a perfectly well-defined spectral problem for the cylinder Hamiltonian.\footnote{The analogous problem is exactly solvable for the $T \bar T$ deformation on $S^1 \times \mathbb{R}$ \cite{Cavaglia:2016oda}.} What's more, we can use the familiar spin-chain language to represent states \cite{Minahan:2002ve}, so that the question of integrability of our spectral problem
becomes the question of integrability of a  specific spin-chain Hamiltonian, which can be analyzed by standard techniques.

The first nice surprise is that the ``rigid'' subalgebra of the original $\mathfrak{psu}(2, 2|4)$ superconformal symmetry that is preserved by the irrelevant deformation is 
$\mathfrak{psu}(2|2) \times \mathfrak{psu}(2|2) \ltimes \mathbb{R}^2$. This is precisely the algebra that plays a crucial role in integrability of ${\cal N}=4$ SYM theory. In the  standard SYM case,
the breaking to this subalgebra occurs spontaneously, by the choice of the BMN vacuum for the spin-chain Hamiltonian. In our case, the breaking is explicit, by the change in the spin-chain Hamiltonian induced by the irrelevant deformation. Nevertheless, the algebraic consequences are the same. Beisert's analysis~\cite{Beisert:2005tm} of the magnon dispersion relation 
and of the  2 $\to$ 2 magnon scattering matrix using  centrally extended $\mathfrak{psu}(2|2)$
 goes through with no essential modification. In particular, the magnon excitations are gapless, and
 symmetry arguments are sufficient to fix uniquely their 2 $\to$ 2 scattering, up to an undetermined dynamical phase. It follows that just as in the standard ${\cal N}=4$ case, the Yang-Baxter equation for magnon scattering is automatically satisfied. These are very nice structural features, but
 integrability is not guaranteed: one  still needs to check factorization of  the S-matrix for multi-magnon processes, which is a dynamical question.
 
 The second nice surprise regards the construction of the classical deformed Lagrangian on $S^3 \times \mathbb{R}$. Writing down a supersymmetric Lagrangian 
 on a curved manifold is a priori a non-trivial task, see, e.g.,~\cite{Festuccia:2011ws}. As our deformation is irrelevant, it entails an infinite set of corrections with increasing powers of $h /\ell^4$,
 already at the classical level, just  to preserve supersymmetry. Remarkably, following \cite{Berkovits:1993hx, Evans:1994np, Pestun:2007rz}, we have found a novel off-shell formalism, precisely tailored for the symmetries of our background. 
 We are able to realize off-shell eight of the sixteen preserved supercharges and to write a close form expression for the classical supersymmetric action on  $S^3 \times \mathbb{R}$ in the presence
 of the irrelevant deformation. Of course, this is just the classical action. As we have emphasized, the quantum theory suffers  from the ambiguities parametrized by the finite values of the counterterms.
 Our hope is that there is at least one choice of counterterms for which integrability is preserved.

 To test this idea, we proceed to study the spin-chain Hamiltonian  in planar perturbation theory.  We focus on a closed subsector, analogous to the $\mathfrak{su}(2|3)$ sector of ${\cal N}=4$ SYM;
the deformation breaks the symmetry down to  $\mathfrak{su}(2|2) \ltimes \mathbb{R}$.
  The spin-chain Hamiltonian admits a double series  expansion,  in the 't Hooft coupling $\lambda$ and in the deformation parameter $h /\ell^4$. They play a similar role: higher powers of both parameters correlate to the spin-chain interactions becoming less and less short-range. It is then convenient to organize
 the expansion in terms of an effective ``loop order'' which measures the non-locality of the chain (one-loop is nearest neighbor, two-loops is next-to-nearest neighbor, etc.).  The most efficient way to fix the spin chain Hamiltonian to low orders in this loop expansion is to leverage symmetry, following the purely algebraic approach of  \cite{Beisert:2003ys}. 
  A tedious but straightforward generalization of Beisert's method
yields the spin-chain Hamiltonian  up to  two loops. When the dust settles, one finds that the smaller symmetry of the deformed theory 
 is still sufficient to fix the result uniquely  (to this order): there is an accidental symmetry enhancement to $\mathfrak{su}(2|3)$ and one recovers the integrable Hamiltonian of ${\cal N}=4$ SYM.  A two-loop calculation is thus  not sufficient for a dynamical test of integrability. A three-loop calculation is beyond the scope of this paper, but we make some structural observations on a conjectural 
integrable long-range spin chain that would preserve the smaller symmetry of the deformed theory and be truly distinct from the ${\cal N}=4$ spin chain. The upshot is that we see no conceptual obstacles
in constructing such a long-range spin chain.

It would be of great interest to find an  integrable irrelevant deformation of ${\cal N}=4$ SYM on $S^3 \times \mathbb{R}$. Such a deformation would be worth studying in its own right, even if  it is not obviously related
to the irrelevant deformation of the  theory on $\mathbb{R}^4$ and its proposed duality with the full D3 brane background, which  was one of the initial motivations for this work.
 A plausible  holographic interpretation of  maximally supersymmetric irrelevant deformations on $S^3 \times \mathbb{R}$  is that they are instead dual to the bubbling geometries of 
  Lin, Lunin and Maldacena \cite{Lin:2004nb}. We offer some brief comments about this interpretation.

  \medskip
 
 The remainder of the paper is organized as follows. In section \ref{classical_sec} we explain how to construct a classical action for the deformed theory that preserves the rigid superalgebra on $S^3 \times \mathbb{R}$. 
 In section \ref{spinchain_sec} we study the spectral problem. We review the structural features guaranteed by the triply centrally extended $\mathfrak{psu}(2|2)$ symmetry;
 we fix the spin-chain Hamiltonian in the $\mathfrak{su}(2|2) \ltimes \mathbb{R}$ subsector up to two loops (accidentally recovering the ${\cal N}=4$ result); and we comment on the possible integrability of the all-orders long-range spin chain. In section \ref{holography_sec} we discuss a possible holographic interpretation. 
We conclude in section \ref{conclusions_sec} with a brief discussion and a list of open questions. Many technical details are relegated to six appendices.

\section{Deformation on \texorpdfstring{$S^3\times \mathbb{R}$}{S³xR} and off-shell classical action}


\label{classical_sec}

Thanks to the conformal flatness of $S^3\times \mathbb R$, one can straightforwardly place $\mathcal N=4$ super Yang-Mills theory on this geometry by performing a Weyl transformation. The symmetry algebra of the curved-space theory will still contain the full $\mathfrak{psu}(2,2|4)$ superconformal algebra. Its Lagrangian density is obtained by covariantizing the flat-space one and adding conformal masses to the scalar fields, {i.e.},
\begin{equation}\label{LYMonshell}
\mathcal L_{\text{YM}} = \tr\left(\frac{1}{2} F_{MN}F^{MN} + \bar \Psi \Gamma^M D_M \Psi + \frac{R}{6} \Phi_I \Phi^I\right)\;,
\end{equation}
where $R$ is the scalar curvature of space-time, which for $S^3\times \mathbb R$ is given by $R= \frac{6}{\ell^2}$ with $\ell$ the radius of $S^3$. As is often useful, we have used ten-dimensional notations: the indices $M,N,\ldots$ take the values $0,1,2,\ldots 9$ and can be split into the four-dimensional space-time indices $\mu,\nu=0,1,2,3$ and the six-dimensional R-symmetry indices $I,J,\ldots = 4,5,6,7,8,9$. The temporal direction lies along $\mathbb R$. The ten-dimensional gauge field $A_M$ thus contains the four-dimensional connection $A_\mu$ and six scalars $\Phi_I$. Furthermore, $\Psi$ is a 32-component spinor satisfying the standard Majorana-Weyl condition and $\Gamma_M$ are ten-dimensional gamma-matrices. All fields are valued in the adjoint representation of the gauge group. We choose the generators anti-Hermitian and thus the covariant derivative acts as $D_M = \partial_M + [A_M,\cdot]$. We refer the reader to appendix \ref{notations} for a summary of our notations and conventions. The Lagrangian density in \eqref{LYMonshell} is invariant (up to total covariant derivatives) under the on-shell transformation rules
\begin{equation}\label{susyvariations}
\delta A_M = \bar \Psi \Gamma_M \varepsilon\;,\qquad \delta \Psi = -\frac{1}{2} F_{MN} \Gamma^{MN}\varepsilon - \frac{1}{2} \Phi_I \Gamma^{\mu I} D_\mu \varepsilon\;.
\end{equation}
Here $\varepsilon$ is a conformal Killing spinor -- it satisfies the equation $D_\mu \varepsilon = \frac{1}{4}\Gamma_\mu \Gamma^\nu D_\nu \varepsilon$ -- and is of Majorana-Weyl type. See appendix \ref{app:N=4onS3S1} for more details.

Our aim is to add to this curved-space Lagrangian the irrelevant deformation. Upon turning on the deformation, we can only hope to preserve an algebra of ``rigid'' ({i.e.}, non-conformal) superisometries. This algebra is a subalgebra of $\mathfrak{psu}(2,2|4)$ and a supersymmetrization of the algebra of (bosonic) isometries of $S^3\times\mathbb R$, {i.e.},
\begin{equation}\label{isometryS3R}
\mathfrak{su}(2)_{\mathcal L} \times \mathfrak{su}(2)_{\mathcal R} \times \mathbb R_\mathcal{H}\;.
\end{equation}
Indeed, the three-sphere can be viewed as the group-manifold of $SU(2)$, and the two $SU(2)$ factors in the isometry group $SO(4) \simeq SU(2)_{\mathcal L} \times SU(2)_{\mathcal R}$ correspond to the left and right action of $SU(2)$ on this group-manifold. Furthermore, $\mathbb R_\mathcal{H}$ denotes translations along the temporal direction generated by $\mathcal H$. Various supersymmetrizations of the algebra \eqref{isometryS3R}, preserving half the number of supersymmetry charges, {i.e.}, sixteen, exist. We will argue in subsection \ref{subsec:flatspace} that the only feasible choice for our purposes is
\begin{equation}\label{superisometries}
\mathfrak{psu}(2|2) \times \mathfrak{psu}(2|2) \ltimes \mathbb{R}^2\;,
\end{equation}
whose bosonic subalgebra contains, in addition to the algebra of isometries in \eqref{isometryS3R}, the R-symmetry subalgebra
\begin{equation}\label{Rsymmsubalg}
\mathfrak{su}(2)_a \times \mathfrak{su}(2)_{\dot a} \times \mathfrak{u}(1)_{\mathcal J} \subset \mathfrak{so}(6)_R\;.
\end{equation}
The supercharges retained in the ``rigid'' algebra of \eqref{superisometries} are selected by the requirement that they commute with $\mathcal H - \mathcal J$. We denote them as
\begin{equation}\label{supercharges}
\mathcal Q^{a\alpha}\;,\qquad \mathcal Q^{\dagger}_{a\alpha}\;, \qquad \tilde{\mathcal Q}_{\dot a\dot\alpha}\;,\qquad \tilde{\mathcal Q}^{\dagger \dot a\dot \alpha}\;,
\end{equation}
where $\alpha (\dot\alpha)$ is an $\mathfrak{su}(2)_{\mathcal L}$($\mathfrak{su}(2)_{\mathcal R}$) index. Note that $\mathcal Q, \tilde{\mathcal Q}$ carry $U(1)_{\mathcal J}$ charge $+1/2$, while their conjugates have $U(1)_{\mathcal J}$ charge $-1/2$. In appendix \ref{app:rigidsub} we provide more details from the viewpoint of the associated Killing spinors. We remark that the generator $\mathcal H - \mathcal J$ is the common central extension of the $\mathfrak{psu}(2|2)$ algebras. Associated with the breaking of the full $SO(6)_R$ R-symmetry as in \eqref{Rsymmsubalg}, we often rename the six scalars $\Phi_I$ into the standard fields $Z,\bar Z$, and $\phi_{a\dot a}$. The latter we often also endow with an $SO(4)_R$ vector index, {i.e.}, $\phi_j$. The $U(1)_{\mathcal J}$ charge assignments are $\mathcal J(Z)=+1,\mathcal J(\bar Z)=-1,\mathcal J(\phi_{a\dot a})=0$.

The flat-space irrelevant term $h\, \mathcal O_8$ must be accompanied by curvature corrections to indeed preserve the ``rigid'' algebra of \eqref{superisometries}. Before discussing these corrections, let us first consider the flat-space deformation $h\, \mathcal O_8$ itself. One can build this operator, for example, by acting with the eight Poincar\'e supercharges of positive $U(1)_{\mathcal J}$ charge on the bottom component $\tr\ \bar Z^4$. These flat-space Poincar\'e variations can be described by the transformation rules in \eqref{susyvariations} for suitably specialized constant Killing spinors.\footnote{The full flat-space conformal Killing spinors are given by $\varepsilon(x)=\varepsilon_{s} + x^\mu \Gamma_\mu \varepsilon_c$, where $\varepsilon_{s}$ and $\varepsilon_{c}$ are constant spinors describing Poincar\'e and special conformal supercharges respectively.} Of course, one antisymmetrizes the supersymmetry variations to avoid generating anticommutators. The resulting operator, however, is only supersymmetric when using the field equations of the undeformed theory, as the supersymmetry variations \eqref{susyvariations} only close upon using these equations of motion. In other words, the newly constructed action of the deformed theory is only supersymmetric to order $h$. To improve on the situation, one can attempt to add higher order terms in $h$ to both the deformation term and the supersymmetry variations, which, because the deformation is irrelevant and thus the parameter $h$ has negative mass dimension, will generate an infinite series of corrections to both. This is clearly an undesirable state of affairs, the more so because we will have to add $1/\ell$ corrections to every single one of these correction terms. Fortunately, following \cite{Berkovits:1993hx,Evans:1994np}, we found an elegant off-shell formalism that produces in one fell swoop the full deformed action including its curvature corrections.

$\mathcal N=4$ super Yang-Mills theory requires seven bosonic auxiliary fields $K^\ell$ to offset the mismatch between fermionic and bosonic off-shell degrees of freedom. We split the seven auxiliary fields as $7=3+4$, and declare that the set of three fields transforms as a spatial vector on $S^3$ while the group of four fields transform as a vector of $SO(4)_R = SU(2)_a \times SU(2)_{\dot a}$. In other words, $K^\ell \rightarrow (K^{\hat{\mu}},K^j)$. As explained in more detail in appendix \ref{app:off-shell}, we can then realize off-shell, while preserving the full isometries of $S^3\times \mathbb R$, the eight supercharges of the superisometry algebra \eqref{superisometries} with positive $U(1)_{\mathcal J}$ charge, {i.e.}, $\mathcal Q^{a\alpha}$ and $\tilde{\mathcal Q}_{\dot a\dot\alpha}$. What's more, the realization is linear. It is given concretely by
\begin{align}\label{offshelltransformation1bismt}
\delta_{\mathcal Q,\tilde{\mathcal Q}}\, A_M &= \bar \Psi \Gamma_M \varepsilon\;,\\\label{offshelltransformation2bismt}
\delta_{\mathcal Q,\tilde{\mathcal Q}}\, \Psi &= -\frac{1}{2} F_{MN} \Gamma^{MN}\varepsilon - \frac{i}{\ell} \Phi^I \Gamma_{I0}\varepsilon - i K^{\hat m} \Gamma_{\hat{m}0} \varepsilon - i K^{j} \Gamma_{j0} \varepsilon\;,\\\label{offshelltransformation3bismt}
\delta_{\mathcal Q,\tilde{\mathcal Q}}\,  K_{\hat m} &= -i  D_M \bar\Psi \Gamma^M \Gamma_{\hat m 0}\varepsilon \;,\\\label{offshelltransformation4bismt}
\delta_{\mathcal Q,\tilde{\mathcal Q}}\,  K_{j} &= -i  D_M \bar\Psi \Gamma^M \Gamma_{j 0}\varepsilon \;.
\end{align}
As emphasized in the notation `` $\delta_{\mathcal Q,\tilde{\mathcal Q}}$ '', these transformation rules are only valid for the selected eight supercharges $\mathcal Q^{a\alpha}$ and $\tilde{\mathcal Q}_{\dot a\dot\alpha}$ of $\mathfrak{psu}(2|2) \times \mathfrak{psu}(2|2) \ltimes \mathbb{R}^2$. One can verify that the off-shell algebra of these transformations closes onto gauge transformations with gauge-parameter the field $Z$. The Yang-Mills Lagrangian density invariant under these off-shell transformation rules is
\begin{equation}\label{offshellYM}
\mathcal L_{\text{YM}}^{\text{off-shell}} = \tr\left(\frac{1}{2} F_{MN}F^{MN} + \bar \Psi \Gamma^M D_M \Psi + \frac{1}{\ell^2} \Phi_I \Phi^I - K^{\hat \mu}K_{\hat \mu} - K^j K_j\right)\;.
\end{equation}

The construction of the irrelevant deformation itself is facilitated by setting up a curved-space superspace formalism.\footnote{One could consider attempting to set up an $\mathcal N=1$ superspace formalism based on the results of \cite{Festuccia:2011ws}. Constructing the appropriate $\mathcal N=1$ curved space D-term is, however, challenging. For some flat-space, Abelian results in this direction, see for example \cite{Tseytlin:1999dj}.} Such formalism is possible because the off-shell transformation rules \eqref{offshelltransformation1bismt}-\eqref{offshelltransformation4bismt} are linear in the Killing spinor (and don't contain derivatives acting on the Killing spinor anymore). We introduce Grassmann-odd coordinates $\theta_{a\alpha}, \tilde{\theta}^{\dot a\dot\alpha}$ and declare that the supercharges are realized as
\begin{equation}\label{superchargesOnSuperfield}
\mathcal Q^{a\alpha} \longleftrightarrow \frac{\partial}{\partial \theta_{a\alpha}} - i \epsilon^{ab}\epsilon^{\alpha\beta}\theta_{b\beta}[ Z , \cdot]\;, \qquad \tilde{\mathcal Q}_{\dot a\dot\alpha} \longleftrightarrow \frac{\partial}{\partial \tilde\theta^{\dot a\dot\alpha}}  - i \epsilon_{\dot a\dot b}\epsilon_{\dot \alpha \dot \beta}\tilde\theta^{\dot b\dot \beta}[ Z , \cdot] \;.
\end{equation}
This realization correctly implements the supersymmetry algebra. On this superspace, we can define a superfield $\bar Z(\theta_{a\alpha}, \tilde\theta^{\dot a\dot\alpha})$ with bottom component $\bar Z$:
\begin{align}
\bar Z(\theta_{a\alpha}, \tilde\theta^{\dot a\dot\alpha})&= \bar Z - 2i \epsilon^{ab}\epsilon^{\alpha\beta}\Psi_{-a\alpha} \, \theta_{b\beta} - 2i \epsilon_{\dot a \dot b}\epsilon_{\dot\alpha\dot \beta} \Psi_{-}^{\dot a\dot\alpha} \, \tilde\theta^{\dot b\dot\beta} \nn\\
&\quad+i \epsilon^{ab}\epsilon^{\alpha\beta}\left(\frac{1}{2}F^{mn} (\sigma_{mn})_\alpha^{\phantom{\alpha}\gamma} \delta_a^c+ \frac{1}{2}[\phi^i,\phi^j] (\sigma_{ij})_a^{\phantom{\alpha}c}\delta_\alpha^\gamma +iK^{\hat m}(\sigma_{\hat m 0})_\alpha^{\phantom{\alpha}\gamma}\delta_a^c \right)\theta_{c\gamma}\theta_{b\beta}\nn\\
&\quad+i \epsilon_{\dot a\dot b}\epsilon_{\dot\alpha\dot \beta} \left( \frac{1}{2}F^{mn} (\bar\sigma_{mn})^{\dot\alpha}_{\phantom{\alpha}\dot\gamma} \delta_{\dot c}^{\dot a} + \frac{1}{2}[\phi^i,\phi^j] (\bar\sigma_{ij})^{\dot a}_{\phantom{a}\dot c} \delta_{\dot\gamma}^{\dot\alpha}  +iK^{\hat m} (\bar\sigma_{\hat m 0})^{\dot\alpha}_{\phantom{\alpha}\dot\gamma} \delta_{\dot c}^{\dot a}  \right) \tilde\theta^{\dot c\dot\gamma} \tilde\theta^{\dot b\dot\beta}\nn\\
&\quad +  2i \epsilon_{\dot c\dot b}\epsilon_{\dot\zeta\dot \beta} \left(D^m \phi_{j} - i\delta^m_0 \left(K_j + \frac{1}{\ell} \phi_j\right)\right)(\bar\sigma^{j})^{\dot c c} (\bar\sigma_{m})^{\dot\zeta \zeta} \theta_{c\zeta}\tilde\theta^{\dot b\dot\beta}\nn\\
&\quad +\ldots \label{Zbarsuperfield}
\end{align}
Here the subscript $-$ on the fermionic fields denotes the (negative) $U(1)_\mathcal J$ charge. The full expression of $\bar Z(\theta_{a\alpha}, \tilde\theta^{\dot a\dot\alpha})$ can be found in appendix \ref{app:Zbarsuperfield}. Its schematic content, omitting commutators, $1/\ell$ corrections, and auxiliary fields, is
\begin{equation}\label{Zbarmultiplet}
\begin{tikzcd}[row sep = scriptsize]
{} & {} &\arrow{dl} \bar Z \arrow{dr} & {} & {}\\
{} & \arrow{dl}  \Psi_{-a\alpha} \arrow{dr} & {}  & \arrow{dl} \Psi_{-}^{\dot a\dot\alpha} \arrow{dr} & {}\\
(F^-)_{\alpha\beta} \arrow{dr} & {} & \arrow{dl} D_m \phi_{a\dot a} \arrow{dr}   & {} & \arrow{dl} (F^+)^{\dot\alpha\dot\beta}\\
{} &   D_{m}\Psi_{+\alpha}^{\dot a} \arrow{dr}  & {} &  \arrow{dl} D_{m}\Psi_{+a}^{\dot \alpha}  & {}\\
{} & {} & D_m D_p Z & {} & {}
\end{tikzcd}
\end{equation}
Note also that
\begin{equation}
\delta_{\mathcal Q,\tilde{\mathcal Q}}\, Z = 0\;.
\end{equation}

With this superfield in hand, we can now easily write down the full off-shell action of the deformed theory:
\begin{equation}\label{deformedaction}
S(g_{\text{YM}}, h) = \frac{1}{g_{\text{YM}}^2} \int_{S^3\times \mathbb R} d^4x \sqrt{|g|}\ \mathcal L_{YM}^{\text{off-shell}}\ + \ h \int_{S^3\times \mathbb R} d^4x \sqrt{|g|} \int d^4\theta d^4\tilde\theta\ \ \tr\, \bar Z(\theta, \tilde\theta)^4 \, + \text{h.c.}
\end{equation} 
This action is manifestly invariant under the eight supercharges $\mathcal Q^{a\alpha}$ and $\tilde{\mathcal Q}_{\dot a\dot\alpha}$. It is straightforward to verify that it is also invariant on-shell under the other eight supercharges provided one turns on a background $U(1)_\mathcal J$ connection $V$ along the temporal direction
\begin{equation}
V = \frac{1}{\ell} dt\;, \qquad D_t = \partial_t - \frac{i}{\ell} \mathcal J + [A_t, \cdot]\;.
\end{equation}
The desired effect of this background connection is that the central element $\mathcal H-\mathcal J$, that makes an appearance in the anticommutators of the supercharges and their daggered counterparts, acts as a temporal derivative. The on-shell invariance under the supercharges $\mathcal Q^\dagger, \tilde{\mathcal Q}^\dagger$ can then be established by realizing that these charges annihilate $\bar Z$ and by observing that when anticommuting the daggered supercharge through the eight supercharges that act on $\tr\, \bar Z^4$, one only encounters combinations of bosonic supercharges whose action, inside the space-time integral, vanishes.

Finally note that upon integrating out the auxiliary fields in \eqref{deformedaction}, one generates an infinite power expansion in $h$.

\subsection{Flat-space limit}\label{subsec:flatspace}
Before analyzing the deformed action of \eqref{deformedaction}, let us provide a complementary point of view on its construction. This alternative viewpoint comes about by considering the decompactification limit of the space $S^3\times \mathbb R$, {i.e.}, $\ell\rightarrow \infty$. Equivalently, we can look at the physics in the tangent space to a point on $S^3\times \mathbb R$. From the Killing spinor equation \eqref{KSequationplus} and its analog for the supercharges of negative $U(1)_{\mathcal J}$ charge, it is straightforward to write down the flat-space expression of the supercharges \eqref{supercharges}. They are
\begin{align}\label{flatspace1}
&\mathcal Q^{a\alpha} \rightarrow Q_+^{a\alpha} + \frac{i}{2\ell} (\bar\sigma_0)^{\dot \alpha \alpha}\, \tilde S_{+\dot \alpha}^{a} \;, \qquad &&\tilde{\mathcal Q}_{\dot a\dot\alpha} \rightarrow \tilde Q_{+\dot a\dot\alpha} + \frac{i}{2\ell} (\sigma_0)_{\alpha \dot \alpha}\, S_{+\dot a}^{\alpha}\;,\\
&\mathcal Q^\dagger_{a\alpha} \rightarrow  \frac{1}{2\ell}S_{- a\alpha} + i (\sigma_0)_{\alpha\dot\alpha}\tilde Q_{-a}^{\dot \alpha}\;, \qquad &&\tilde{\mathcal Q}^{\dagger \dot a\dot\alpha} \rightarrow \frac{1}{2\ell} \tilde S_-^{\dot a\dot\alpha} + i (\bar \sigma_0)^{\dot\alpha\alpha}Q_{-\alpha}^{\dot a}\;.
\end{align}
It is easy to verify that their leading order anticommutators are those of the standard Poincar\'e algebra.

Let us now consider the deformed on-shell (and thus to leading order in $h$) Lagrangian at the point we chose, which we label the origin in the tangent space. It is clear that it is given by
\begin{equation}\label{flatspacedef}
\prod_{a,\alpha} \big(Q_+^{a\alpha} + \frac{i}{2\ell} (\bar\sigma_0)^{\dot \alpha \alpha}\, \tilde S_{+\dot \alpha}^{a}\big)\ \prod_{\dot a,\dot\alpha}\big(\tilde Q_{+\dot a\dot\alpha} + \frac{i}{2\ell} (\sigma_0)_{\alpha \dot \alpha}\, S_{+\dot a}^{\alpha}\big) \ \tr\, \bar Z(0)^4\;.
\end{equation}
Evaluating this expression is straightforward: the action of the Poincar\'e/special conformal supercharges will simply move us up/down in the $\mathbf{105}$ supermultiplet. Its result will take the form 
\begin{equation}\label{expansion1/l}
\mathcal O_8(0) + \frac{\mathcal O_7(0)}{\ell} + \ldots +\frac{\mathcal O_4(0)}{\ell^4}\;,
\end{equation}
where the operators $\mathcal O_n$ can be identified with components of dimension $n$ operators occurring in the $\mathbf{105}$ supermultiplet. What's more, these components are necessarily singlets of the tangent space symmetry subgroup $SO(3)\subset SO(3,1)$ and of the R-symmetry subgroup $SO(4)_R \times U(1)_{\mathcal J}$, as these are the symmetries preserved by the combinations of supercharges that are acting in \eqref{flatspacedef}.

For example, the operator $\mathcal O_4$ can be easily constructed. Recalling that the adjoint representation of $SU(4)$ decomposes as $\mathbf{15}\rightarrow (\mathbf 3,\mathbf 1)_0 \oplus (\mathbf 1,\mathbf 3)_0 \oplus (\mathbf 2,\mathbf 2)_{+2} \oplus (\mathbf 2,\mathbf 2)_{-2}$ under the subgroup $SU(2)_a\times SU(2)_{\dot a} \times U(1)_{\mathcal J}$, it is clear that the nonzero anticommutators of the Poincar\'e and special conformal supercharges appearing in \eqref{flatspace1} result in R-symmetry generators in the $(\mathbf 2,\mathbf 2)_{+2}$. At order $\ell^{-4}$ in \eqref{flatspacedef}, we then find
\begin{equation}\label{orderellm4_flat}
\mathcal O_4 = \frac{1}{16} (\det R_{(\mathbf 2,\mathbf 2)_{+2}})^2\ \tr\, \bar Z^4 = \frac{3}{2} \text{Str}(3 Z^2\bar Z^2 - 6 Z\bar Z \phi_j \phi_j + \phi_i\phi_i\phi_j\phi_j)\;.
\end{equation}
where we used the symmetrized trace. This analysis can be continued and produces the full leading order deformed action, to which we will return in the next subsection.

Note that we can now also argue that the only feasible supersymmetrization of \eqref{isometryS3R} is the one in \eqref{superisometries}. The argument starts by making the trivial observation that $\mathcal O_8$ is always the top-component of the $\mathbf{105}$ supermultiplet. Now, for any supersymmetrization of \eqref{isometryS3R}, we can consider the flat-space limit as above. Acting with the flat-space limit of any of the supercharges, which necessarily involves a special conformal supercharge, on $\mathcal O_8$, it is clear that the special conformal supercharge will not annihilate $\mathcal O_8$, but rather it takes us one level down in the supermultiplet. In other words, $\mathcal O_8$ is not supersymmetric by itself and we need to include a bosonic operator $\mathcal O_7$ such that its Poincar\'e variation cancels the $S$-variation of $\mathcal O_8$. Moreover, this operator should be a singlet under the preserved R-symmetry and under the $SO(3)\subset SO(3,1)$ preserved spatial subgroup. The special conformal variation of $\mathcal O_7$ is however non-zero, demanding the addition of $\mathcal O_6$ and so forth. This process keeps going until we reach an operator that is annihilated by all special conformal supercharges appearing in the flat-space limit of the supercharges. A democratic supersymmetrization of the two $SU(2)$ factors of \eqref{isometryS3R} as in \eqref{superisometries} can be easily seen to be a valid option. Indeed, the $\mathbf{105}$-supermultiplet contains candidate operators $\mathcal O_i$ singlet under both the preserved R-symmetry of \eqref{Rsymmsubalg} and the spatial subgroup. However, trying to supersymmetrize \eqref{isometryS3R} into $\mathfrak{su}(2|4)\times \mathfrak{su}(2)$ fails already when looking for an $SU(4)_R$ singlet operator $\mathcal O_7$, while $\mathcal O_8$ is acted on by all special conformal supercharges. Similarly, aiming to preserve $\mathfrak{su}(2|3)\times \mathfrak{su}(2|1)$ does not work either. There does not exist a candidate operator $\mathcal O_4$ because the representation $\mathbf {105}$ does not contain a singlet when decomposed into $SU(3)\times U(1)$ representations, neither does the $\mathbf{105}$ supermultiplet contain a candidate for $\mathcal O_5$. For $\mathcal O_6$ there is a single candidate originating from an $SO(3,1)$ singlet operator transforming in the $SU(4)_R$ representation $(2,0,2)=\mathbf {84}$. However, it is easy to convince oneself that it is not annihilated by all $S$ appearing in the flat-space limit of the supercharges.\footnote{More in detail, the operator is given by $\epsilon^{\alpha\beta}\epsilon_{\dot\alpha\dot\beta} S^{(\mathcal I}_{\alpha}S^{\mathcal J)}_{\beta} \tilde S^{\dot\alpha}_{(\mathcal K}\tilde S^{\dot\beta}_{\mathcal L)} \mathcal O_8(0)$, where $\mathcal I, \mathcal J,\ldots$ are $SU(4)_R$ indices, and further symmetrizations are needed to produce the representation $\mathbf {84}$ and to avoid the occurrence of special conformal generators. The singlet under $SU(3)\times U(1)$ is simply $\epsilon^{\alpha\beta}\epsilon_{\dot\alpha\dot\beta} S^{4}_{\alpha}S^{4}_{\beta} \tilde S^{\dot\alpha}_{4}\tilde S^{\dot\beta}_{4} \mathcal O_8(0)$. Acting on this expression with any of the charges $\tilde S_{\mathcal I=1,2,3}^{\dot\gamma}$ or $S^{\mathcal I=1,2,3}_{\gamma}$ clearly does not vanish: they can be anticommuted through the $S$'s with impunity, but act nontrivially on $\mathcal O_8$.}

\subsection{Leading order deformed action}
Let us analyze the action \eqref{deformedaction} in some more detail to leading order in $h$. To this order, the integration over the auxiliary fields is easy to perform. They appear purely quadratically in the Yang-Mills action (see \eqref{offshellYM}), and thus their field equation is $K = \mathcal O(h)$. We can thus effectively set them to zero in the leading order deformed action. The computation of the superspace integral in \eqref{deformedaction} is technical but straightforward. Alternatively, to this order in $h$, we can just as well evaluate \eqref{flatspacedef}. Either way, the resulting Lagrangian density takes the form
\begin{equation}\label{leadinddeformedL}
\mathcal L(g_{\text{YM}}, h) = \frac{1}{g_{\text{YM}}^2} \mathcal L_{YM}\ + \ h \ \left[\mathcal O_8 + \frac{\mathcal O_7}{\ell} + \ldots +\frac{\mathcal O_4}{\ell^4}\right] + \mathcal O(h^2)\;,
\end{equation} 
where, as we have already argued above, the operators $\mathcal O_n$ can be identified with components of dimension $n$ operators occurring in the $\mathbf{105}$ supermultiplet. Recall that these components necessarily are singlets of the tangent space symmetry subgroup $SO(3)\subset SO(3,1)$ and of the R-symmetry subgroup $SO(4)_R \times U(1)_{\mathcal J}$. The former requirement implies that the operator of which we are considering particular components has Lorentz spins $j_1=j_2$.

For example, $\mathcal O_8$ is the top-component of the multiplet. Considering for simplicity only the leading term in $g_{\text{YM}}$, we find 
\begin{align}
\mathcal O_8 = \tr\Big[ & F_{mn} F^{np}F_{pq} F^{qm} -\frac{1}{4} (F_{mn} F^{nm})^2 + 4\big(F_{mp}F^{np} - \tfrac{1}{4} F_{pq}F^{pq}\delta_{m}^{n}\big) D^m \Phi_I D_n \Phi_I  \notag \\
& -(D_m\Phi_I)(D^m\Phi_I)(D_n\Phi_J)(D^n\Phi_J) +2 (D_m\Phi_I)(D^m\Phi_J)(D_n\Phi_I)(D^n\Phi_J) \notag \\
& + (\text{terms with fermions}) \notag \\
& + (\text{total derivatives and terms proportional to field equations}) \notag \\
& + \mathcal O(g_{\text{YM}})\Big]\label{O8expression}
\end{align}
Note that the equations of motion contain terms of higher order in both $h$ and $\ell^{-1}$. The former are part of the $\mathcal O(h^2)$ corrections in \eqref{leadinddeformedL}, while the latter (at order $h^0$) have additional terms at order $\ell^{-2}$. The result in \eqref{O8expression} is precisely the leading term of the $\mathcal N=4$ supersymmetric DBI action, see for example equation (4.28) of \cite{Tseytlin:1999dj}. More generally, one may be tempted to speculate that the $\ell\rightarrow \infty$ limit of the complete action $S(g_{\text{YM}}, h)$ is the full flat-space non-abelian DBI action. However, as was also the case for the all-order DBI-deformation in \cite{Chang:2014nwa}, there is no compelling reason for such a speculation -- one needs a principle to fix  D-term ambiguities.

To identify the operator $\mathcal O_7$, capturing the $\ell^{-1}$-correction, we first recall that the $Q^3\tilde Q^3 $ descendant of the $\mathbf{105}$ supermultiplet is a dimension seven operator that transforms as a spatial vector and in the adjoint representation of the $SU(4)_R$ symmetry. See, for example, entry $\mathcal O^{(17)}_0$ of table 1 of \cite{DHoker:1999mic}, the relevant part of which we reproduce in table \ref{105multiplet}. Note that our setup guarantees that the $Y$-symmetry quantum number is zero for all $\mathcal O_n$. The operator $\mathcal O_7$ can now be recognized (up to total derivatives and terms proportional to equations of motion) as the component that is a singlet of $SO(3)\subset SO(3,1)$ and $SO(4)_R \times U(1)_{\mathcal J}\subset SU(4)_R$.

\begin{table}[t]
\centering
\begin{tabular}{ l | c | c | c | c }
operator & desc & dim & $(j_1,j_2)$ & $SU(4)$ rep \\
\hline\hline
$\mathcal O^{(0)}_4 \sim \tr\, \Phi^4$ & $-$ & 4 & $(0,0)$ & $(0,4,0) = \mathbf{105}$ \\
$\mathcal O^{(3)}_2 \sim \tr\, \lambda\tilde\lambda\Phi^2$ & $Q\tilde Q$ & 5 & $(\tfrac{1}{2},\tfrac{1}{2})$ & $(1,2,1) = \mathbf{175}$ \\
$\mathcal O^{(9)}_2 \sim \tr\, F_+F_-\Phi^2$ & $Q^2\tilde Q^2$ & 6 & $(1,1)$ & $(0,2,0) = \mathbf{20'}$ \\
$\mathcal O^{(12)}_0 \sim \tr\, \lambda\lambda\tilde\lambda\tilde\lambda$ & $Q^2\tilde Q^2$ & 6 & $(0,0)$ & $(2,0,2) = \mathbf{84}$ \\
$\mathcal O^{(17)}_0 \sim \tr\, F_+F_-\lambda\tilde\lambda$ & $Q^3\tilde Q^3$ & 7 & $(\tfrac{1}{2},\tfrac{1}{2})$ & $(1,0,1) = \mathbf{15}$ \\
$\mathcal O^{(20)}_0 \sim \tr\, F_+^2F_-^2$ & $Q^4\tilde Q^4$ & 8 & $(0,0)$ & $(0,0,0) = \mathbf{1}$ 
\end{tabular}
\caption{\label{105multiplet} Operators in $\mathbf{105}$ supermultiplet with bonus symmetry quantum number $Y=0$ and rotational quantum numbers $j_1=j_2$.}
\end{table}

This type of analysis can be continued. The operator $\mathcal O_6$ is a particular linear combination of the $Q^2\tilde Q^2$ descendants in the $\mathbf{105}$ supermultiplet that contain singlets under the decomposition into the preserved spatial and R-symmetry group. There are two candidates, namely, in the notation of table 1 of \cite{DHoker:1999mic}, $\mathcal O_0^{(12)}$, which transforms as a spatial singlet in the $\mathbf {84}$ of $SU(4)_R$, and $\mathcal O_2^{(9)}$, which is a spatial symmetric traceless tensor that transforms in the  $\mathbf {20'}$ of $SU(4)_R$. Similarly, $\mathcal O_5$ is given by the appropriate component of $\mathcal O_2^{(3)}$. Finally, we arrive at $\mathcal O_4$, which can only be the unique singlet of the $\mathbf{105}$ representation under $SO(4)_R \times U(1)_{\mathcal J}$ and was reported in \eqref{orderellm4_flat} already.

\section{Spectral problem and spin chains}

\label{spinchain_sec}

When quantizing a relativistic theory on $S^3 \times \mathbb{R}$, we can regard the time translation generator ${\mathcal{H}}$ as an operator acting on the Hilbert space of states on $S^3$. In the absence of the deformation, this corresponds to the standard dilatation operator and time translations on the cylinder are equivalent to dilatations on $\mathbb{R}^4$. Nevertheless, time translations on the cylinder remain a symmetry even when the irrelevant deformation is turned on  and therefore it is a physically meaningful question to enquire about the spectrum of ${\mathcal{H}}$. This is the problem we will be addressing in this section.

The main difficulty is to resolve the mixing of states under renormalization. The large $N$ limit simplifies the analysis in that it automatically establishes  the standard parametrically large suppression of the mixing of single with multiple traces. By single and multiple-trace states, we mean the states that correspond to single and multiple-trace operators respectively under the state-operator map when the irrelevant is turned off and we get back to the original conformal $\mathcal{N}=4$ SYM.
From now on we will consider only the single trace sector and define the corresponding states as
\beq \label{statedef}
|\chi_1 \dots \chi_L \rangle \xleftrightarrow{\text{state-operator map}} \Tr \left( \chi_1 \dots \chi_L \right) \text{ when }h=0\,,
\eeq
for a given set of elementary fields $\{\chi_1,\dots, \chi_m\}$.

\subsection{Dispersion relation and two-body S-matrix}
One of the most appealing properties of the deformation under study is its underlying  
$\mathfrak{psu}(2|2) \times \mathfrak{psu}(2|2) \ltimes \mathbb{R}^2$ symmetry algebra. Both $\mathfrak{psu}(2|2) $ copies share the common central extension $\mathcal{H}-\mathcal{J}$ where $\mathcal{H}$ and $\mathcal{J}$ are the time-translation and the $\mathfrak{u}(1)_{\mathcal J}$ generators, respectively. As in  \cite{Beisert:2005tm}, we enlarge the algebra by two additional unphysical central charges which allow for nontrivial representations of $\mathfrak{psu}(2|2)$ depending on a free continuous parameter related to the coupling constants.
It turns out that this triply extended symmetry by itself is sufficiently constraining to guarantee a number of nonperturbative properties in the spectrum, some of those are required for integrability as we will see below. 

To make the symmetry manifest, it is customary to write the elementary fields transforming in the bifundamental of $\mathfrak{psu}(2|2)^2$ in the following form,
\beq
\{ \Phi_{a\dot{a}}, \Psi_{a}{}^{\dot{\alpha}},\Psi^{\alpha}{}_{\dot{a}},\mathcal{D}^{\alpha \dot{\alpha}} \},\,\,\,\,a,\dot{a}=1,2\;\;\text{and}\;\;\alpha,\dot{\alpha}=\pm\,.
\eeq
where dotted and undotted indices belong to a fundamental representation of each copy of $\mathfrak{psu}(2|2)$, respectively.
These are the excitations or magnons to be inserted on top of the vacuum state. The vacuum state is in turn constructed out of the $U(1)_{\mathcal J}$ charged field $Z$ as
\beq
|Z^{J}\rangle\,.
\eeq
 Contrary to the $\mathcal{N}=4$ SYM case, the magnons are not Goldstones as this vacuum does not break any global symmetry: the $U(1)_{\mathcal J}$ group is singled out  from the get-go. 
 We will be considering infinite chains where the charge $J$ is taken to be  larger than the range of the interactions (perturbatively the range of the interactions is finite at a given loop order), but nevertheless the states are required to be periodic. 
 
 Spin chains which contain a centrally extended $\mathfrak{psu}(2|2)$ symmetry are rather special as shown in \cite{Beisert:2005tm}, in that  the single and double excitation dynamics is strongly constrained. For example, we can run precisely the same arguments to show that the energy of a single magnon is fixed to be 
 \beq \label{dispersion}
 E(p)-J =\frac{1}{2}\sqrt{1+ 16\, \alpha(g^2,h/\ell^4) \sin^2 \left( \frac{p}{2}\right) }\,,
 \eeq
where $\alpha(g^2,h/\ell^4)$ is an unfixed function of the coupling constants and $p$ is the magnon momenta which is  quantized from the periodicity condition. As a consequence, we observe that the excitations remain massless despite not being interpreted as Goldstone particles.
Another important direct consequence of this symmetry concerns the two-body scattering matrix. Up to a global phase and coupling redefinition, the S-matrix is otherwise fixed and its explicit expression can be found in \cite{Beisert:2005tm}. It turns out to be a solution of the Yang-Baxter equation and although this fact \textit{per se} is not a sufficient condition for the integrability of the model, it is nevertheless a good hint. In order to probe integrability, we have to determine the Hamiltonian itself and investigate whether it is a conserved charge in involution with a larger set of commuting charges. This is will be the goal of the next section.

\subsection{Perturbative \texorpdfstring{$\mathfrak{su}(2|2)\ltimes \mathbb{R}$}{su(2|2)xR}  Hamiltonian }\label{perturbation}
In the large $N$ limit, 
the problem of determining the Hamiltonian on $S^3\times \mathbb{R}$  is technically analogous to the one of determining the dilatation operator in planar  $\mathcal{N}=4$ SYM.  We will be looking at the particular $\mathfrak{su}(2|2)\ltimes \mathbb{R}$ closed subsector, whose field content is 
\beq
\phi_{a} \equiv \Phi_{a \dot{1}}\,,\;\;\; \psi^{\alpha} \equiv \Psi^{\alpha}{}_{\dot{1}}\,,\;\;\; Z\,.
\eeq
where $a=1,2$ and $\alpha=\pm$\,.  The generators of $\mathfrak{su}(2|2)\ltimes \mathbb{R}$ satisfy the following commutation relations
\begin{align}
 \label{algebra}
\{\mathcal{Q}^{\dagger}_{a\alpha} , \mathcal{Q}^{b \beta}\} &=  \delta^{b}_{a}\, \mathcal{L}_{\alpha}{}^{\beta} + \delta_{\alpha}^{\beta}\, \mathcal{R}_{a}{}^{b} +\frac{1}{2} \delta^{b}_{a}\, \delta^{\beta}_{\alpha} \left( \mathcal{H}-\mathcal{J} \right) 
\\
\{\mathcal{Q}^{a \alpha},\mathcal{Q}^{b \beta} \} &= \epsilon^{ab}\, \epsilon^{\alpha \beta} \, \mathcal{P}\,,\quad\qquad\quad\,  \{\mathcal{Q}^{\dagger}_{a\alpha},\mathcal{Q}^{\dagger}_{b \beta}\} = \epsilon_{ab} \, \epsilon_{\alpha \beta}\, \mathcal{K}
\\
[\mathcal{R}_{a}{}^{b}, \mathcal{G}^{c}] &= -\delta^{c}_a\, \mathcal{G}^{b}+\frac{1}{2}\,\delta^{b}_{a} \, \mathcal{G}^{c}\,,\quad\quad  [\mathcal{R}_{a}{}^{b}, \mathcal{G}_{c}] = \delta^b_{c}\, \mathcal{G}_{a}-\frac{1}{2}\,\delta^{b}_{a} \, \mathcal{G}_{c}
\\
[\mathcal{L}_{\alpha}{}^{\beta}, \mathcal{G}^{\gamma}]&=\delta^{\gamma}_{\alpha}\, \mathcal{G}^{\beta}-\frac{1}{2}\,\delta^{\beta}_{\alpha}\,\mathcal{G}^{\gamma}\,, \quad\quad\, [\mathcal{L}_{\alpha}{}^{\beta}, \mathcal{G}_{\gamma}]=-\delta^{\beta}_{\gamma}\, \mathcal{G}^{\beta}+\frac{1}{2}\,\delta^{\beta}_{\alpha}\,\mathcal{G}_{\gamma}
\\
[\mathcal{H-\mathcal{J}},\mathcal{Q}^{a \beta}] &= [\mathcal{H-\mathcal{J}},\mathcal{Q}^{\dagger}_{b\alpha}]=0\,,
\end{align}
where $\mathcal{J}$ is the $\mathfrak{u}(1)_{\mathcal{J}}$  generator and $\mathcal{H}$ is the time translation generator, the two extended  central charges (besides $\mathcal{H}-\mathcal{J}$) are $\mathcal{P}$ and $\mathcal{K}$, and $\mathcal{G}$ is a generic generator with an $\mathfrak{su}(2)_{\alpha}$ or $\mathfrak{su}(2)_{a}$ index. 
This sector is the counterpart of the maximally compact subsector $\mathfrak{su}(2|3)$ of $\mathcal{N}=4$ SYM upon identifying $Z$ with $\phi^3$, except that now the symmetry is smaller. We are interested in studying what are the consequences of this symmetry on the perturbative form of the Hamiltonian $\mathcal{H}$. 

Our deformation is hermitian and preserves the spin chain parity. Parity is related to the following $\mathbb{Z}_2$ transformation of the $SU(N)$ generators
\beq \label{parity}
 T^{A}{}_{B}\,  \rightarrow - T^{B}{}_{A}\,.
\eeq
Consequently, we can define a parity operator $P$ that acts on a state as defined in (\ref{statedef}), as
\beq
P | \chi_{1} \dots \chi_m \rangle = (-1)^{m+ n_{f}(n_{f}+1)/2} |\chi_1 \dots \chi_{m} \rangle
\eeq
with $n_f$ being the number of fermions in the state.
The action (\ref{deformedaction}) can be verified 
to be invariant under this symmetry\footnote{The effect of the parity operation amounts to inverting the order of the fields within a trace and introducing a minus sign if the number of fields is odd. Since the action is quartic and fully symmetric under the permutation of the fields, it remains invariant.} and hence we expect the Hamiltonian to inherit this property, namely
\beq
[P,\mathcal{H}] = 0\,.
\eeq  
We assume that under the RG flow this classical symmetry is preserved and we will impose parity symmetry at every order in the perturbative expansion. 

We will now follow closely the strategy outlined in \cite{Beisert:2003ys}: in short, we allow for quantum corrections to the symmetry generators  and impose that order by order in the loop expansion, the algebraic relations (\ref{algebra}) are preserved. Compatibility with the algebra will then impose constraints on these corrections. 

A practical way of parametrizing the invariant tensor structures from which the generators are made of is by means of the symbols introduced in \cite{Beisert:2003ys},
\beq \label{symbols}
\PTerm{A_1\ldots A_n}{B_1\ldots B_m}
\eeq
which act on a spin chain state by replacing the sequence of fields $A_1 \dots A_n$ by the sequence $B_1\dots B_m$, or kill the state in case such sequence does not show up (see \cite{Beisert:2003ys} for more details). A generic generator $\mathcal{G}$  admits an expansion 
in standard perturbation theory
\beq \label{couplexp}
\mathcal{G}(\alpha) = \sum_{m=0} \kappa^{m} \mathcal{G}_{m},
\eeq
where the parameter $\kappa$ is related to the coupling constants of the theory as we now explain. Generically, we will have a double expansion in  $g^2$ and $h$ but we now argue that they are on the same footing and therefore a term of order $\mathcal{O}(g^{2 j} (h/\ell^4)^i = \mathcal{O}(\kappa^{j+i})$
contributes to the $j+i$ loop order in the perturbative expansion. 
\begin{figure}[t]
\begin{center}
\includegraphics[clip,height=4cm]{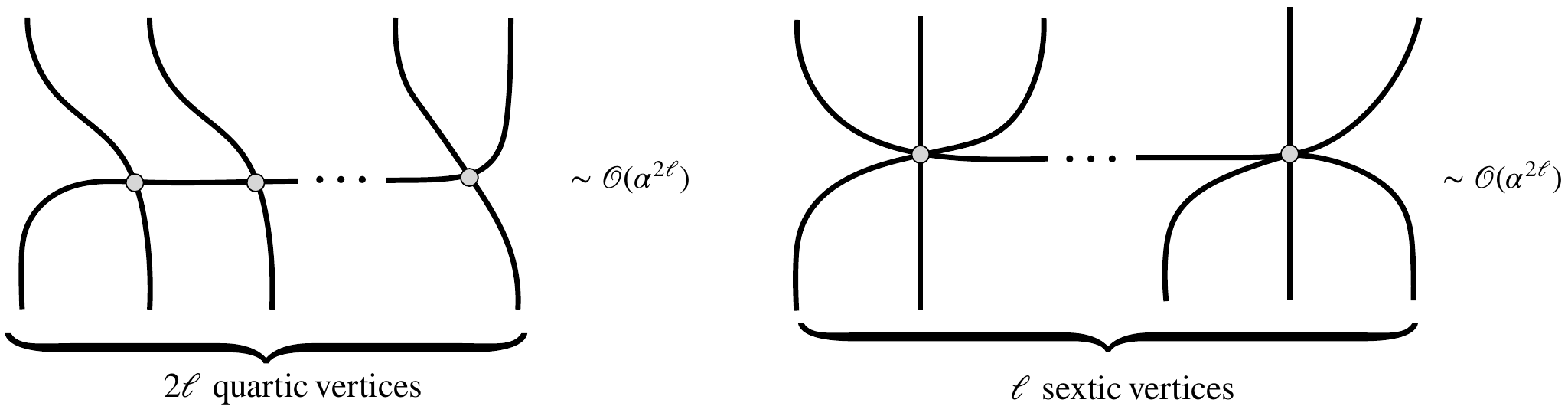} 
\end{center}
\vspace{-0.7cm}
\caption{ The left graphs contains $2\ell$ quartic vertices (hence it can appear at least at $2 \ell$ loops in the gauge theory) and has range $2 \ell +1$. The graph on the right contains $\ell$ sextic vertices but has the same range and it appears at the same loop order as the left graph. 
} 
\label{fig:graphs0}
\end{figure}
Recall that the classical off-shell Lagrangian (\ref{offshellYM}) includes seven auxiliary bosonic scalar fields $K^{l}$, which have to be integrated out and this will produce higher order terms in the coupling $h/\ell^4$. These fields appear quadratically in the original Super Yang-Mills action in the form
\beq
\mathcal L_{\text{YM}}^{\text{off-shell}}=\tr \left( \dots - K^{\hat \mu}K_{\hat \mu} - K^j K_j\right),
\eeq 
and at the leading order in the deformation $h/\ell^4$, all fields appear quartically including the auxiliary ones as can be explicitly checked in the example (\ref{O8expression}). This means that their equations of motion will be schematically of the form
\beq
K = h \, X^{3}+\dots
\eeq
where $X$ can be any of the fields. This in turn has the consequence that a term of order $j$ in the coupling $h$  will contain a vertex with at most  $2j+2$ elementary fields.  For example, for $j=2$, the term in front of $h^2$  contains sextic scalar vertices.

The range of planar interactions from such new vertices is then the same as in the original $\mathcal{N}=4$ SYM, see figure \ref{fig:graphs0} for an example. We then conclude that, as in $\mathcal{N}=4$ SYM, the symbols (\ref{symbols}) should contain at most $k+2$ fields at order $k$ in the expansion in the coupling $\kappa$, see (\ref{couplexp}).

\subsubsection{Leading order}
At leading order, the most general ansatz for all generators compatible with $\mathfrak{su}(2)_{a}\times \mathfrak{su}(2)_{\alpha}\times \mathfrak{u}(1)_{\mathcal{J}}$ symmetry is
\beq \label{treegens}
\begin{aligned}
\mathcal{R}_{b}{}^{a} &= c_1 \PTerm{a}{b}+c_2 \delta^{a}_{b} \PTerm{c}{c}\\
\mathcal{L}_{\beta}{}^{\alpha} &= c_3 \PTerm{\alpha}{\beta}+c_4 \delta^{\alpha}_{\beta} \PTerm{\gamma}{\gamma}\\
\mathcal{Q}^{a \alpha} &= c_5 \PTerm{a}{\alpha}\\
\mathcal{Q}^{\dagger}_{a\alpha} &= c_6 \PTerm{\alpha}{a}\\
\mathcal{H}_0 &=  c_7 \PTerm{a}{a}+c_{8} \PTerm{\alpha}{\alpha}+ c_{9}\PTerm{Z}{Z} \\
\mathcal{J}&=  c_{10} \PTerm{Z}{Z} + c_{11}\PTerm{\alpha}{\alpha} \,.
\end{aligned}
\eeq
The only nontrivial solution that realizes the algebra is given by
\beq
c_1=c_3=c_7= 1,\quad c_2=c_4=-c_{11}= -\frac{1}{2},\quad c_6= e^{-i \beta_1}= \frac{1}{c_5},\quad c_8= \frac{3}{2},\quad c_{10}= c_9 = e^{i \delta_1}\,.
\eeq
The unfixed parameters $\beta_1$ and $\delta_1$ are simply a manifestation of the similarity transformations that can be performed on the generators $\mathcal{G}(\alpha)$ without affecting the algebra. They can be interpreted as the freedom in rescaling the normalization of the bosonic and fermionic fields. From now on, we will fix the constant $\delta_1=0$ and leave $\beta_{1}$ arbitrary.

\subsubsection{Higher orders}
Ultimately we will be interested in determining the spectrum of the Hamiltonian. A useful observation, largely explored in \cite{Kim:2002if,Beisert:2003ys}, is that one can always perform a similarity transformation $\mathcal{H}(\alpha) \rightarrow U(\alpha) \, \mathcal{H}(\alpha) \, U(\alpha)^{-1}$, such that $\mathcal{H}(\alpha)$ can be diagonalized on a basis with definite classical energy. In other words, in such a basis we have $[\mathcal{H}(\alpha),\mathcal{H}_0]=0$, regardless of the field theory being conformal or not. This means that the mixing problem is significantly simplified because the classical energy $E_0$ is preserved in the transitions induced by $\mathcal{H}(\alpha)$. 

Using this observation, it is immediate to write  at one-loop the most general structure of $\mathcal{H}_2$ compatible with the symmetries and conservation of classical energy,
\beq
\begin{aligned} \label{ham2}
\mathcal{H}_2 &=  d_1 \PTerm{a b}{ab}+ d_2 \PTerm{a \beta}{a \beta} + d_2' \PTerm{\alpha b}{\alpha b}+d_3 \PTerm{\alpha \beta}{\alpha \beta}+d_4 \PTerm{ab}{ba}+d_5 \PTerm{a \beta}{\beta a}\\
&+d_5'\PTerm{\alpha b}{b \alpha} + d_6 \PTerm{\alpha \beta}{\beta \alpha} +d_{00} \PTerm{Z Z}{Z Z}+d_{01} \PTerm{Z \beta}{Z \beta}+d_{02} \PTerm{\beta Z}{\beta Z}+d_{03} \PTerm{\beta Z}{Z \beta}\\
&+d_{04} \PTerm{Z \beta}{\beta Z}+d_{05} \PTerm{Z b}{ Z b}+d_{06} \PTerm{ b Z}{  b Z}+d_{07} \PTerm{ Z b}{  b Z}+d_{08} \PTerm{ b Z}{ Z b}\,.
\end{aligned}
\eeq
Imposing parity invariance gives 6 constraints,
\beq
d_2'=d_{2},\quad d_5'=d_{5},\quad d_{02}= d_{01},\quad d_{04}=d_{03},\quad d_{06}=d_{05},\quad d_{08}=d_{07}
\eeq
which leaves 11 parameters to be fixed.
Out of those, 10 turn out to be fixed by imposing the  algebraic relations up to third order leaving one single overall constant $d_1 \equiv \alpha_1^{2}$, which can be interpreted as the freedom of redefining the coupling and cannot be settled through this algebraic procedure. The resulting Hamiltonian at this order is then given by
\beq
\begin{aligned}
\mathcal{H}_2 =   \alpha_1^2& \Bigl(\PTerm{a b}{ab}+ \PTerm{a \beta}{a \beta} + \PTerm{\alpha b}{\alpha b}+ \PTerm{\alpha \beta}{\alpha \beta} -\PTerm{ab}{ba}- \PTerm{a \beta}{\beta a}-\PTerm{\alpha b}{b \alpha}+\PTerm{\alpha \beta}{\beta \alpha}\\
& +\PTerm{Z \beta}{Z \beta}+ \PTerm{\beta Z}{\beta Z}+\PTerm{Z b}{ Z b}+\PTerm{ b Z}{  b Z}- \PTerm{Z \beta}{\beta Z}-\PTerm{ Z b}{  b Z}- \PTerm{ b Z}{ Z b}-\PTerm{\beta Z}{Z \beta} \Bigr)\,.
\end{aligned}
\eeq
We then conclude that at first order the symmetry gets enhanced and we get back to the $\mathfrak{su}(2|3)$ Hamiltonian of $\mathcal{N}=4$ SYM up to the coupling definition. This means at this order in this particular subsector the theory is integrable.

We can proceed to higher loops in a systematic way: at each order, the generators are made out of linear combinations of the structures in (\ref{symbols}) that preserve the classical energy as well as the R-charge, and the number of fields involved should be compatible with the loop order as discussed in section \ref{perturbation}. 
Then imposing parity, hermiticity and closure of the algebra puts constraints on the allowed space of parameters. 
Upon eliminating additional unphysical parameters related to similarity transformations, we find for the perturbative corrections to the time translation generator $\mathcal{H}$ up to order $\alpha^4$ (see appendix \ref{pertham} for some more details):

\begin{scriptsize}
\begin{align}
\mathcal{H}_0\,&=\, \PTerm{a}{a}+\frac{3}{2} \PTerm{\alpha}{\alpha}+ \PTerm{Z}{Z}\displaybreak[0]\\
\mathcal{H}_2\,&=\,  \alpha_1^2\, \Bigl(\PTerm{a b}{ab}+ \PTerm{a \beta}{a \beta} + \PTerm{\alpha b}{\alpha b}+ \PTerm{\alpha \beta}{\alpha \beta} -\PTerm{ab}{ba}- \PTerm{a \beta}{\beta a}-\PTerm{\alpha b}{b \alpha}+\PTerm{\alpha \beta}{\beta \alpha} +\PTerm{Z \beta}{Z \beta}+ \PTerm{\beta Z}{\beta Z}+\PTerm{Z b}{ Z b}\nn\\
&+\PTerm{ b Z}{  b Z}- \PTerm{Z \beta}{\beta Z}-\PTerm{ Z b}{  b Z}- \PTerm{ b Z}{ Z b}-\PTerm{\beta Z}{Z \beta} \Bigr)
\displaybreak[0]\\
\mathcal{H}_3 \, &= \,\frac{\alpha_1^3\, e^{i \beta_2}}{\sqrt{2}}  \epsilon_{a b} \epsilon_{\alpha \beta } \PTerm{\alpha \beta }{Z ab}+\frac{\alpha_1^3 e^{i \beta_2}}{\sqrt{2}} \epsilon_{a b} \epsilon _{\alpha \beta } \PTerm{\alpha \beta }{a b Z}+\frac{\alpha_1^3 e^{-i \beta_2}}{\sqrt{2}} \epsilon_{a b} \epsilon_{\alpha ,\beta } \PTerm{a Z b}{\alpha \beta}
\nn\\
&-\frac{\alpha_1^3 e^{i \beta_2}}{\sqrt{2}} \epsilon_{a b} \epsilon_{\alpha ,\beta } \PTerm{\alpha \beta }{a Z b}-\frac{\alpha_1^3 e^{-i \beta_2}}{\sqrt{2}} \epsilon_{a b} \epsilon_{\alpha ,\beta } \PTerm{Z a b}{\alpha \beta}-\frac{\alpha_1^3 e^{-i \beta_2}}{\sqrt{2}} \epsilon_{a b} \epsilon_{\alpha ,\beta } \PTerm{a b Z}{\alpha \beta}
\displaybreak[0]\\
\mathcal{H}_4\, &=\, \left(-2 \alpha _1^4+2 \alpha _3 \alpha _1-\alpha _2\right) \PTerm{a b c}{a b c}+\left(\frac{3 \alpha _1^4}{2}-\alpha _1 \alpha _3\right) (\PTerm{a b c}{a c b}+\PTerm{a b c}{b a c})+\left(-2 \alpha _1^4+2 \alpha _3 \alpha _1-4 \alpha _2\right) \PTerm{a Z c}{a Z c}-4 \alpha _2 \PTerm{a Z c}{c Z a}
\displaybreak[0]\nn\\
&+\left(\frac{\alpha _1^4}{2}+2 \alpha _3 \alpha _1-6 \alpha _2\right) \PTerm{a \beta  c}{a \beta  c}+\left(-\frac{3}{2} \alpha _1^4+\alpha _3 \alpha _1-\frac{3 \alpha _2}{2}\right) (\PTerm{a b Z}{a Z b}+\PTerm{a Z c}{a c Z}+\PTerm{a Z c}{Z a c}+\PTerm{Z b c}{b Z c})
\displaybreak[0]\nn\\
&+\left(\frac{\alpha _1^4}{2}+\frac{\alpha _2}{2}\right) (\PTerm{a b Z}{b Z a}+\PTerm{a Z c}{c a Z}+\PTerm{a Z c}{Z c a}+\PTerm{Z b c}{c Z b})+\left(2 \alpha _1 \alpha _3-2 \alpha _1^4\right) (\PTerm{a b Z}{a b Z}+\PTerm{Z b c}{Z b c})
\displaybreak[0]\nn\\
&+\left(\frac{3 \alpha _1^4}{2}-\alpha _3 \alpha _1+2 \alpha _2\right) (\PTerm{a b Z}{b a Z}+\PTerm{Z b c}{Z c b})+\left(-2 \alpha _1^4+2 \alpha _3 \alpha _1+\alpha _2\right) \PTerm{Z b Z}{Z b Z}+\left(\alpha _1 \alpha _3-\frac{\alpha _1^4}{2}\right) (\PTerm{a Z Z}{a Z Z}+\PTerm{Z Z c}{Z Z c})
\displaybreak[0]\nn\\
&+\frac{\alpha _1^4}{2} (\PTerm{a Z Z}{Z Z a}+\PTerm{Z b \gamma }{\gamma  b Z}+\PTerm{Z Z c}{c Z Z}+\PTerm{Z Z \gamma }{\gamma  Z Z})+\left(\frac{\alpha _1^4}{2}+2 \alpha _3 \alpha _1-2 \alpha _2\right) (\PTerm{a \beta  Z}{a \beta  Z}+\PTerm{Z \beta  c}{Z \beta  c})+\left(\frac{\alpha _1^4}{2}+2 \alpha _3 \alpha _1+\alpha _2\right) \PTerm{Z \beta  Z}{Z \beta  Z}
\displaybreak[0]\nn\\
&+\left(-\frac{5}{4} \alpha _1^4+\alpha _3 \alpha _1-\alpha _2\right) (\PTerm{a b \gamma }{a \gamma  b}+\PTerm{a \beta  c}{a c \beta }+\PTerm{a \beta  c}{\beta  a c}+\PTerm{\alpha  b c}{b \alpha  c})+\left(\alpha _2-\frac{\alpha _1^4}{2}\right) (\PTerm{a b \gamma }{\gamma  b a}+\PTerm{\alpha  b c}{c b \alpha })
\displaybreak[0]\nn\\
&+\left(2 \alpha _1 \alpha _3-\frac{11 \alpha _1^4}{4}\right) (\PTerm{a b \gamma }{a b \gamma }+\PTerm{\alpha  b c}{\alpha  b c})+\left(-\frac{5}{4} \alpha _1^4+\alpha _3 \alpha _1-2 \alpha _2\right) (\PTerm{a \beta  Z}{\beta  a Z}+\PTerm{Z b \gamma }{Z \gamma  b}+\PTerm{Z \beta  c}{Z c \beta }+\PTerm{\alpha  b Z}{b \alpha  Z})
\displaybreak[0]\nn\\
&+\left(\frac{\alpha _1^4}{4}-\alpha _2\right) (\PTerm{a \beta  Z}{Z a \beta }+\PTerm{Z b \gamma }{b \gamma  Z}+\PTerm{Z \beta  c}{\beta  c Z}+\PTerm{\alpha  b Z}{Z \alpha  b})+\left(-\frac{11}{4}  \alpha _1^4+2 \alpha _3 \alpha _1+2 \alpha _2\right) (\PTerm{Z b \gamma }{Z b \gamma }+\PTerm{\alpha  b Z}{\alpha  b Z})
\displaybreak[0]\nn\\
&+\left(-\alpha _1^4+\alpha _3 \alpha _1-\alpha _2\right) (\PTerm{a b \gamma }{b a \gamma }+\PTerm{a \beta  \gamma }{\beta  a \gamma }+\PTerm{\alpha  b c}{\alpha  c b}+\PTerm{\alpha  b \gamma }{b \alpha  \gamma })-\frac{\alpha _1^4}{4} (\PTerm{a b \gamma }{b \gamma  a}+\PTerm{a \beta  c}{c a \beta }+\PTerm{a \beta  c}{\beta  c a}+\PTerm{a \beta  \gamma }{\gamma  a \beta }+\PTerm{\alpha  b c}{c \alpha  b}+\PTerm{\alpha  b \gamma }{b \gamma  \alpha })
\displaybreak[0]\nn\\
&+\left(-4 \alpha _1^4+2 \alpha _3 \alpha _1+2 \alpha _2\right) \PTerm{\alpha  b \gamma }{\alpha  b \gamma }+\left(\frac{\alpha _1^4}{2}-\alpha _2\right) (\PTerm{a b c}{b c a}+\PTerm{a b c}{c a b}+\PTerm{a b Z}{Z a b}+\PTerm{Z b c}{b c Z}+\PTerm{\alpha  b \gamma }{\gamma  b \alpha })
\displaybreak[0]\nn\\
&+\left(-\frac{\alpha _1^4}{2}-4 \alpha _2\right) (\PTerm{a Z \gamma }{\gamma  Z a}+\PTerm{\alpha  Z c}{c Z \alpha })+\left(\frac{\alpha _2}{2}\right) (\PTerm{a Z \gamma }{\gamma  a Z}+\PTerm{Z b \gamma }{\gamma  Z b}+\PTerm{\alpha  b Z}{b Z \alpha }+\PTerm{\alpha  Z c}{Z c \alpha })
\displaybreak[0]\nn\\
&+\left(-\frac{5}{4} \alpha _1^4+\alpha _3 \alpha _1-\frac{3 \alpha _2}{2}\right) (\PTerm{a Z \gamma }{a \gamma  Z}+\PTerm{a \beta  Z}{a Z \beta }+\PTerm{Z \beta  c}{\beta  Z c}+\PTerm{\alpha  Z c}{Z \alpha  c})+\left(-\frac{11}{4} \alpha _1^4+2 \alpha _3 \alpha _1-4 \alpha _2\right) (\PTerm{a Z \gamma }{a Z \gamma }+\PTerm{\alpha  Z c}{\alpha  Z c})
\displaybreak[0]\nn\\
&-\frac{\alpha _1^4}{2} (\PTerm{\alpha  b Z}{Z b \alpha }+\PTerm{\alpha  Z Z}{Z Z \alpha })+\left(\alpha _1^4-\alpha _1 \alpha _3\right) (\PTerm{a Z Z}{Z a Z}+\PTerm{Z b Z}{b Z Z}+\PTerm{Z b Z}{Z Z b}+\PTerm{Z Z c}{Z c Z}+\PTerm{Z Z \gamma }{Z \gamma  Z}+\PTerm{Z \beta  Z}{Z Z \beta }+\PTerm{Z \beta  Z}{\beta  Z Z}+\PTerm{\alpha  Z Z}{Z \alpha  Z})
\displaybreak[0]\nn\\
&+\left(\alpha _1 \alpha _3-\frac{7 \alpha _1^4}{4}\right) (\PTerm{Z Z \gamma }{Z Z \gamma }+\PTerm{\alpha  Z Z}{\alpha  Z Z})+\left(-\frac{\alpha _1^4}{4}-\frac{\alpha _2}{2}\right) (\PTerm{a Z \gamma }{Z \gamma  a}+\PTerm{a \beta  Z}{\beta  Z a}+\PTerm{Z \beta  c}{c Z \beta }+\PTerm{Z \beta  \gamma }{\gamma  Z \beta }+\PTerm{\alpha  Z c}{c \alpha  Z}+\PTerm{\alpha  Z \gamma }{Z \gamma  \alpha })
\displaybreak[0]\nn\\
&+\left(-4 \alpha _1^4+2 \alpha _3 \alpha _1-4 \alpha _2\right) \PTerm{\alpha  Z \gamma }{\alpha  Z \gamma }+\left(\frac{\alpha _1^4}{2}+4 \alpha _2\right) \PTerm{\alpha  Z \gamma }{\gamma  Z \alpha }-\alpha _2 (\PTerm{a \beta  \gamma }{\gamma  \beta  a}+\PTerm{\alpha  \beta  c}{c \beta  \alpha })+\left(\alpha _1^4-\alpha _3 \alpha _1+\alpha _2\right) (\PTerm{\alpha  b \gamma }{\alpha  \gamma  b}+\PTerm{\alpha  \beta  c}{\alpha  c \beta })
\displaybreak[0]\nn\\
&+\left(2 \alpha _1 \alpha _3-4 \alpha _2\right) (\PTerm{a \beta  \gamma }{a \beta  \gamma }+\PTerm{\alpha  \beta  c}{\alpha  \beta  c})+\frac{\alpha _1^4}{4} (\PTerm{\alpha  b \gamma }{\gamma  \alpha  b}+\PTerm{\alpha  \beta  c}{\beta  c \alpha })+\left(-\frac{7}{4}  \alpha _1^4+\alpha _3 \alpha _1-\alpha _2\right) (\PTerm{a \beta  \gamma }{a \gamma  \beta }+\PTerm{\alpha  \beta  c}{\beta  \alpha  c})
\displaybreak[0]\nn\\
&+\alpha _2 (\PTerm{a \beta  c}{c \beta  a}+\PTerm{Z \beta  \gamma }{\beta  \gamma  Z}+\PTerm{\alpha  \beta  Z}{Z \alpha  \beta })\displaybreak[0]\nn\\
&+\left(\alpha _1^4-\alpha _3 \alpha _1+\frac{3 \alpha _2}{2}\right) (\PTerm{a Z \gamma }{Z a \gamma }+\PTerm{Z b \gamma }{b Z \gamma }+\PTerm{Z \beta  \gamma }{\beta  Z \gamma }+\PTerm{\alpha  b Z}{\alpha  Z b}+\PTerm{\alpha  Z c}{\alpha  c Z}+\PTerm{\alpha  Z \gamma }{Z \alpha  \gamma }+\PTerm{\alpha  Z \gamma }{\alpha  \gamma  Z}+\PTerm{\alpha  \beta  Z}{\alpha  Z \beta })
\displaybreak[0]\nn\\
&+2 \alpha _1 \alpha _3 (\PTerm{Z \beta  \gamma }{Z \beta  \gamma }+\PTerm{\alpha  \beta  Z}{\alpha  \beta  Z})+\left(\frac{\alpha _1^4}{4}+\frac{\alpha _2}{2}\right) (\PTerm{\alpha  Z \gamma }{\gamma  \alpha  Z}+\PTerm{\alpha  \beta  Z}{\beta  Z \alpha })+\left(-\frac{7}{4}  \alpha _1^4+\alpha _3 \alpha _1-2 \alpha _2\right) (\PTerm{Z \beta  \gamma }{Z \gamma  \beta }+\PTerm{\alpha  \beta  Z}{\beta  \alpha  Z})
\displaybreak[0]\nn\\
&+\left(2 \alpha _1 \alpha _3-\frac{\alpha _1^4}{2}\right) \PTerm{\alpha  \beta  \gamma }{\alpha  \beta  \gamma }+\left(-\frac{7}{4}  \alpha _1^4+\alpha _3 \alpha _1-3 \alpha _2\right) (\PTerm{\alpha  \beta  \gamma }{\alpha  \gamma  \beta }+\PTerm{\alpha  \beta  \gamma }{\beta  \alpha  \gamma })+2 \alpha _2 (\PTerm{\alpha  \beta  \gamma }{\beta  \gamma  \alpha }+\PTerm{\alpha  \beta  \gamma }{\gamma  \alpha  \beta })-3 \alpha _2 \PTerm{\alpha  \beta  \gamma }{\gamma  \beta  \alpha } 
\end{align}
\end{scriptsize}%
Similarly, for the perturbative corrections to the SUSY generators $\mathcal{Q}$ and $\mathcal{S}$ up to order $\alpha^2$, we find
\begin{scriptsize}
\begin{align}
\left(\mathcal{Q}_0\right)^{a \alpha}\, &= \,e^{i \beta_1} \PTerm{a}{\alpha}
\\
\left(\mathcal{Q}_1\right)^{a \alpha}\,&=\,   \frac{\alpha_1 e^{i \beta _1+i \beta _2}}{\sqrt{2}} \epsilon ^{a b} \epsilon _{\alpha  \beta } \PTerm{\beta }{b Z}-\frac{\alpha_1 e^{i \beta _1+i \beta _2}}{\sqrt{2}} \epsilon ^{a b} \epsilon _{\alpha \beta }  \PTerm{\beta }{Z b}
\\
\left(\mathcal{Q}_2\right)^{a  \alpha}\,&=\,   -\frac{\alpha _1^2}{4}  e^{i \beta _1} \left( \PTerm{ Z a}{ Z \alpha }+  \PTerm{ Z a}{\alpha   Z}+  \PTerm{a  Z}{ Z \alpha }-  \PTerm{a  Z}{\alpha  Z}-  \PTerm{a \beta }{\beta  \alpha }+  \PTerm{\beta  a}{\alpha  \beta }+  \PTerm{a b}{b \alpha }-  \PTerm{a b}{\alpha b}-  \PTerm{b a}{b \alpha }+ \PTerm{b a}{\alpha b} \right)\\
\left(\mathcal{Q}^{\dagger}_0\right)_{a \alpha}\, &= \,e^{-i \beta_1} \PTerm{\alpha}{a}
\\
\left(\mathcal{Q}^{\dagger}_1\right)_{a \alpha}\,&=\,   \frac{\alpha_1 e^{-i \beta _1-i \beta _2}}{\sqrt{2}} \epsilon_{a b} \epsilon ^{\alpha  \beta } \PTerm{b Z}{\beta }-\frac{\alpha_1 e^{-i \beta _1-i \beta _2}}{\sqrt{2}} \epsilon_{a b} \epsilon^{\alpha \beta }  \PTerm{Z b}{\beta }
\\
\left(\mathcal{Q}^{\dagger}_2\right)_{a \alpha}\,&=\,   -\frac{\alpha _1^2}{4}  e^{-i \beta _1} \left( \PTerm{ Z \alpha }{ Z a}+  \PTerm{\alpha   Z}{ Z a}+  \PTerm{ Z \alpha }{a  Z}-  \PTerm{\alpha  Z}{a  Z}-  \PTerm{\beta  \alpha }{a \beta }+  \PTerm{\alpha  \beta }{\beta  a}+  \PTerm{b \alpha }{a b}-  \PTerm{\alpha b}{a b} -  \PTerm{b \alpha }{b a}+ \PTerm{\alpha b}{b a} \right)
\end{align}
\end{scriptsize}%

The outcome of this analysis is the appearance of a new parameter $\alpha_2$ when compared to the $\mathfrak{su}(2|3)$ case. When this parameter is zero, we recover precisely the two loop Hamiltonian of \cite{Beisert:2003ys} in the basis where all the unphysical parameters there are set to zero. The two other unfixed parameters $\alpha_1$ and $\alpha_3$  are associated to coupling redefinition which is of course a symmetry of the algebra: an inspection of the above two-loop Hamiltonian shows that the terms  in front of $\alpha_{3}$ are proportional to $\mathcal{H}_2$,
\beq
\mathcal{H}_4 = \mathcal{H}_{4}\big\rvert_{\alpha_3=0} +2 \alpha_1 \alpha_3 \mathcal{H}_2\,.
\eeq
This can then be interpreted as the freedom of redefining the coupling constant as
$\alpha_1 \kappa \rightarrow  \alpha_1 \kappa + \alpha_3 \kappa^3$.
We need to resort to some additional input to fix them. Here we will use the all-loop dispersion relation or, equivalently, the central extensions of the algebra not considered so far and we will see that in fact the parameter $\alpha_{2}$ will be fixed while the remaining $\alpha_1$ and $\alpha_3$ are left free.

\subsection{Dispersion relation}
We have not yet made use of the two additional central extensions of the algebra besides $\mathcal{H}$. A shortcut to impose the constraints that follow from the additional central extensions  is to consider the magnon dispersion relation. 
As we have briefly described before, the fully centrally extended algebra fixes the form of the all-loop dispersion relation up to the precise definition of the coupling constant. Here, we will compute the perturbative dispersion relation obtained from the previous Hamiltonian and match it with the all-loop prediction (\ref{dispersion}) in order to impose further constraints on the remaining unfixed parameters $\alpha_{1,2,3}$. 
We consider a single-magnon state,
\beq
| \phi_{a}(p) \rangle \equiv \sum_{n=1}^{L} e^{i p n}\, |Z \dots \ \underset{n^{\text{th}}\text{ site }}{\phi_{a}} \dots Z \rangle
\eeq
and a straightforward computation shows that the corresponding energy is given by
\beq
E(p) =\left(-4\, \kappa^2 \alpha _1^2 +8\, \kappa^4 \alpha _1 \alpha _3 \right) \sin ^2\left(\frac{p}{2}\right)+\kappa^4 \alpha _2-8 \kappa^4 \alpha _1^4 \sin ^4\left(\frac{p}{2}\right)\,.
\eeq
Comparison with the expansion of (\ref{dispersion})  sets
\beq
\alpha_2 = 0\,.
\eeq
On the other hand, as we have observed above, $\alpha_{1,3}$ are associated to the precise definition of the coupling constant which cannot be fixed by any of these methods.
The upshot of this analysis is that, as in the one-loop case, we get an enhancement of symmetry leading to the integrable $\mathfrak{su}(2|3)$ spin chain.

\subsection{Long-range spin chains}\label{longrange}
We have shown that the $\mathfrak{su}(2|2)\ltimes \mathbb{R}$ symmetry uniquely fixes the form of the planar time-translation generator to be the same as the integrable planar $\mathcal{N}=4$ SYM dilatation operator up to two loops, apart from the definition of the coupling constant. A natural question is whether there exists integrable $\mathfrak{su}(2|2)\ltimes \mathbb{R}$ long range spin chains which are truly distinct from the planar $\mathcal{N}=4$ SYM one and if so, how large is the corresponding moduli. Clearly the procedure we have employed up to two loops can only take us so far.
Nevertheless, we can systematically address this problem in a smaller and more tractable closed subsector. The simplest is  the $\mathfrak{u}(1)$ sector composed of the elementary fields $\mathcal{Z}$ and $\phi^1$ which corresponds to an XXZ spin chain. Fortunately, such case has been studied in \cite{Beisert:2013voa}. It was found that there is a large class of long-range XXZ spin chains whose moduli can be compactly  encoded in the Bethe ansatz equations for $N$ magnons,
\beq
\exp(i p(u_k) L)=\exp(i \phi L) \prod_{i\neq k}^{N} \exp\left(-2i \theta(u_k,u_i)\right) \frac{\sinh \hbar (u_k-u_i+i) }{\sinh \hbar (u_k-u_i-i)} \,,
\eeq
where the momentum is defined in terms of the rapidity $u$ as
\beq \label{momq}
p(u)=\frac{1}{i} \log \left( \frac{\sinh \hbar (u+i/2)}{\sinh \hbar (u- i/2)}\right) + \sum_{r=2}^{\infty} \gamma_r \, q_{r}(u_k)
\eeq
and $q_{r}(u)$ are the  higher conserved charges given by
\beq
q_{r}(u) = q^{{\rm{NN}}}_{r}(u)+\sum_{s=r+1}^{\infty} \gamma_{r,s}\, q^{{\rm{NN}}}_{s}(u)\,.
\eeq
In this expression, $q^{{\rm{NN}}}_{r}(u)$ are the standard higher conserved charges of the nearest-neighbor XXZ spin chain
\beq
q^{{\rm{NN}}}_{r}(u) = \frac{1}{(r-1)!}\frac{d ^{r-2}}{d u^{r-2}}\left(\frac{i \hbar}{\tanh \left(\hbar \left(u+\frac{i}{2}\right)\right)}-\frac{i \hbar}{\tanh \left(\hbar \left(u-\frac{i}{2}\right)\right)}\right)\,.
\eeq
Finally, the \textit{dressing phase} $\theta$ is given by
\beq \label{theta}
\theta(u_k,u_i) =\sum_{s>r=2}^{\infty} \beta_{r,s} \left(q_{r}(u_k)q_{s}(u_i) -q_{s}(u_k)q_{r}(u_i)  \right)+\sum_{r=2}^{\infty} \eta_{r} \left( q_{r}(u_k) -q_{r}(u_i)   \right)\,,
\eeq
where the parameters $\hbar, \phi, \gamma_r, \gamma_{r,s}, \beta_{r,s},\eta_r$ generally admit an expansion in the coupling constants.
Different choices of these parameters correspond to different integrable models. We note that embedding these spin chains in a larger $\mathfrak{su}(2|2)$ symmetry forces $\hbar =\phi = 0$ to all orders by comparison with the restriction of the full  S-matrix to the $\mathfrak{u}(1)$ subsector. 
Invariance under the centrally extended $\mathfrak{su}(2|2)\ltimes\mathbb{R}^2$ fixes the relation between the momentum $p$ and the rapidity $u$ through the Zhukovski variable $x(u)$,
\beq \label{Zhu}
p(u) = \frac{1}{i} \log \frac{x(u+i/2)}{x(u-i/2)}\,,\,\,\, \text{where } x(u)+\frac{\alpha(g^2,h/\ell^4)}{2\, x(u)} = u\,.
\eeq
This relation sets $\gamma_{2r} = 0$ while the remaining $\gamma_{2r+1}$ can be trivially fixed by matching (\ref{momq}) with (\ref{Zhu}).  Analogously, \cite{Beisert:2013voa} shows that the coefficients $\gamma_{r,s}$ are related to the variable $x(u)$ by
\beq
\frac{1}{x(u)^{r-1}} =\sum_{s=r}^{\infty }\gamma_{r,s} \frac{r-1}{s-1}\frac{1}{u^{s-1}}
\eeq
which can be used to fix $\gamma_{r,s}$. In particular, $\gamma_{r,s}= 0$ for $|r-s|$ odd.
Requiring  a parity invariant Hamiltonian also enforces  $\beta_{r,s} = 0$ for even $r+s$ and $\eta_{2r}=0$.\footnote{This can be seen from the following: at the level of the spin chain, parity conjugation acts by sending the roots $u_{k} \rightarrow -u_{k}$; these conditions are then required to ensure invariance of the Bethe equations.}

 Furthermore, we should impose compatibility with planar Feynman diagrams. Each structure in the Hamiltonian should be consistent with the planar graphs which originates them: at a given fixed loop order, planar graphs have a maximal range and number of interactions, and the corresponding spin chain tensor structures must do so as well. The coefficient $\beta_{r,s}$ leads to interactions of range $s+1$ and generally may arise at order $\mathcal{O}(\kappa^{s-1})$. In the case of $\mathcal{N} = 4$ SYM, it was argued that consistency with the gauge theory vertices delays the appearance of such deformation parameters to $\mathcal{O}(\kappa^{r+s-2})$. 
\begin{figure}[t]
\begin{center}
\includegraphics[clip,height=3cm]{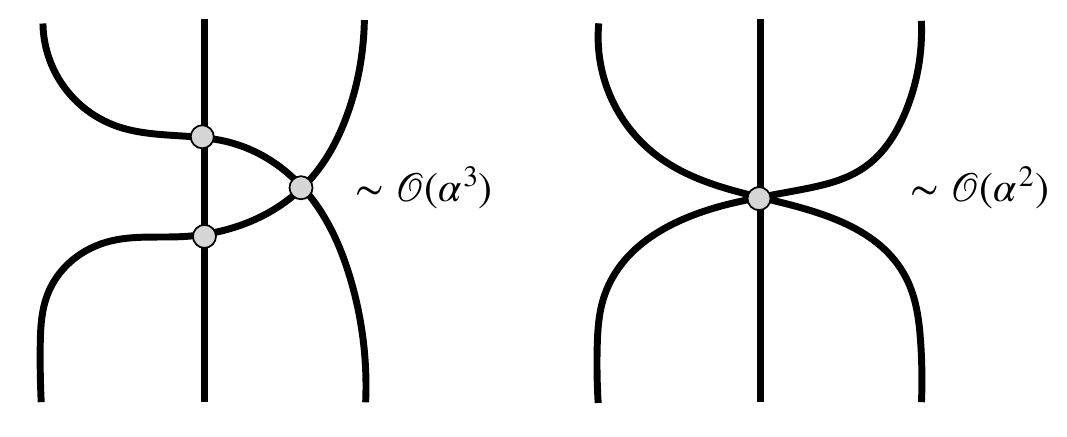} 
\end{center}
\vspace{-0.7cm}
\caption{The left graph illustrates an example of a range three interaction involving three permutations that might appear at three loop order in a gauge theory with at most quartic vertices. On the right graph, the same range three interaction might appear at a lower order in perturbation theory for a gauge theory containing sextic vertices.
} 
\label{fig:graphs}
\end{figure}
In our theory, 
the existence of vertices with $2 r$ legs at $r-1$ loop order allows for the appearance of these coefficients at order
\beq
\beta_{r,s} = \mathcal{O}(\kappa^{s-1})\,\,,
\eeq 
see figure \ref{fig:graphs} for an example.
From the analysis of \cite{Beisert:2013voa}, the parameters $\eta_r$ may arise at order $\mathcal{O}(\kappa^{r-1})$ and are associated to deformations of range $r+1$. For example, it is shown that the first of such parameters, $\eta_2$, arises in front of the following range three structures on the spin chain
\beq
\sigma_{-}\otimes 1 \otimes \sigma_{+}\, \,\,\,\, \text{and} \,\,\,\, \sigma_{+}\otimes 1 \otimes \sigma_{-}\,.
\eeq
These structures could have been generated from a sextic vertex of two loop order, which means that the starting order of these deformations are compatible with the vertices of the gauge theory under study.\footnote{Recall that if a deformation parameter appearing in the Bethe equations is of order $\kappa^r$ then it must appear on the gauge theory Hamiltonian at $r+1$-loop order.} Our explicit two-loop computation of the Hamiltonian shows nevertheless that $\eta_{r}=0$ at order $\kappa$. Beyond this order, there may exist corrections.

A further restriction on the long-range spin chain Hamiltonian comes from crossing symmetry. In integrable models, this property is often used to constraint the S-matrix dressing factor $\theta$. In our case, we will consider the same decomposition of the phase as in $\mathcal{N}=4$ SYM \cite{Arutyunov:2006iu},
\beq \label{dec}
\theta = \chi(x^{+},y^{-})- \chi(x^{+},y^{+})- \chi(x^{-},y^{-})+ \chi(x^{-},y^{+})\,.
\eeq
where $\chi(x,y)$ is an antisymmetric function, $\chi(x,y) = - \chi(y,x)$. Compatibility with (\ref{theta}), requires that $\chi$ should admit an expansion for large $x$ and $y$ of the form
\beq \label{wantedchi}
\chi(u,v) = \sum_{r,s =1}^{\infty} \frac{c_{r,s}}{x^r y^s} + \sum_{r=1} b_r \left( \frac{(y+1/y)}{x^r}-  \frac{(x+1/x)}{y^r} \right)\,.
\eeq
where $c_{r,s}$ and $b_{r}$ are related to the parameters $\beta_{r,s}$ and $\eta_{r}$ in (\ref{theta}).
Being a function of the Zhukovski variables $x$ as in (\ref{Zhu}), we expect to find an analogous type of branch cut structure as in $\mathcal{N}=4$ SYM and therefore, crossing symmetry must be implemented similarly. The crossing equation \cite{Janik:2006dc,Volin:2009uv} must then hold,
\beq \label{cross}
\chi(x,1/y)+\chi(1/x,y)+\chi(1/x,1/y)+\chi(x,y) = \frac{1}{i} \log \frac{\Gamma (1+iu-iv)}{\Gamma(1-iu+iv)}\,,
\eeq
which can be regarded as a Riemann-Hilbert problem for the function $\chi$.
Solving this equation requires specifying the analytic properties and asymptotic conditions of the function $\chi$. It is precisely in this last point where we expect a deviation from the $\mathcal{N}=4$ SYM solution. In that case, the asymptotic limit $x\rightarrow \infty$ is expected to render $\chi(x,y) $ to a constant in such a way that the phase $\theta$ trivializes. This is so as scattering with zero momentum particles is expected to be trivial given their interpretation as a global symmetry transformation of a state with finite momentum excitations. In our context, the global symmetry is broken and hence the scattering with zero momentum particles is not expected to be immaterial. Instead, we already see from (\ref{theta}) that the second term does not become trivial as one of the rapidities, say $u_k$, is sent to infinity.
Therefore one possibility is to consider solutions of the crossing equation of the form
\beq
\chi(u,v) = \chi_{0}(u,v) + \left(y+\frac{1}{y} \right) \chi_{1}(u)- \left(x+\frac{1}{x} \right) \chi_{1}(v)\,.
\eeq
where $\chi_{0}$ is a solution of (\ref{cross}) with the property that $\lim_{x\rightarrow \infty}\chi_{0}(x,y) = \text{constant}$ and $\chi_{1}(x)$ solves the homogeneous equation
\beq \label{cross2}
\chi_1(u+i 0)+\chi_1(u-i 0) = 0\,,\,\,\,\, {\rm{for}}\,\,\, |u|<2 \, \alpha(g^2,h/\ell^4)\,.
\eeq 
In this way, an expansion of $\chi_{0}$  for large $x$ and $y$ will give rise to the first term of (\ref{theta}) whereas $\chi_1$  produces the second term. The equation for $\chi_0$  singles out a unique solution when supplemented with its analytic structure as in $\mathcal{N}=4$ SYM. The equation $(\ref{cross2})$ for $\chi_1$ allows for ample freedom in the choice of the coefficients $\eta_r$ in (\ref{theta})\footnote{For example, any function of the form $\chi_1(u) = \left(g(x)-g(1/x)\right)(f(x)+f(1/x))$ for generic functions $g(x)$ and $f(x)$ solves (\ref{cross2}) and thus crossing.} and further input is needed to fix a particular solution. We then conclude that integrable long-range XXZ spin chains embedded in crossing invariant theories with larger supersymmetry appear to leave a large moduli space of parameters.

We close this section by observing that the phase $\theta$ as given by (\ref{theta}), vanishes for the scattering of two infinite Bethe roots. Given that $1/2$-BPS states are interpreted as Bethe states with all roots at infinity (for which the corresponding energy vanishes), we then expect that any integrable deformed theory with this symmetry preserves the same BPS spectrum as $\mathcal{N}=4$ SYM despite having a smaller global symmetry. This can be regarded as a prediction for a putative holographic dual model.


\section{Comment on holography and LLM geometries}

\label{holography_sec}

Supergravity solutions corresponding to $\mathcal{N}=4$ SYM on $S^{3}\times \mathbb{R}$ have been found by Lin, Lunin and Maldacena (LLM) \cite{Lin:2004nb}. These solutions preserve  $SO(4)\times SO(4)\times \mathbb{R}$ bosonic symmetries and have sixteen supersymmetries, and for this reason they seem to provide a natural setting for an holographic description of the irrelevant deformation studied here. Let us briefly review them. 

\begin{figure}[t]
\begin{center}
\includegraphics[clip,height=7cm]{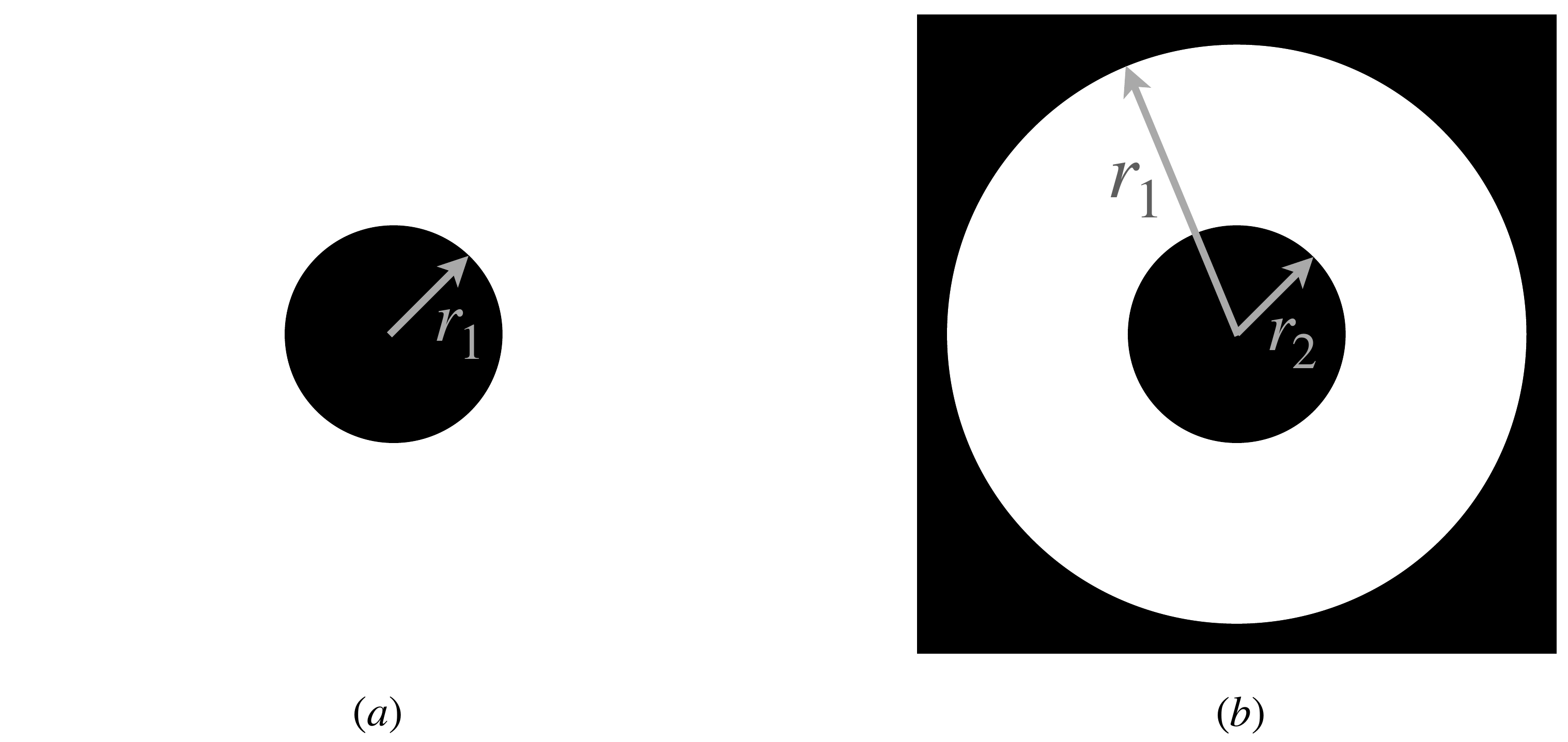} 
\end{center}
\vspace{-0.7cm}
\caption{(a) LLM picture for the $AdS_{5}\times S^{5}$ background with the radius $r_1$ identified with the AdS radius as $r_{1} =R^2_{\rm{AdS}}$. (b) Next-to-simpest LLM geometry with an additional $U(1)$ symmetry. 
} 
\label{fig:LLM}
\end{figure}

These solutions can be uniquely classified by bi-coloring a certain plane (spanned by $r$ and $\varphi$ below) into black and white regions. This bi-colored plane is the result of demanding regularity of the background metric. This  enforces a certain function entering the supergravity solution to take only two possible discrete values in this plane, or equivalently, to make a choice of  black and white colored regions. To make contact with our deformation, we will be considering LLM solutions with an additional $U(1)$ symmetry on the plane. Explicitly, the  background metric is given by
\beq
ds^{2} = -2 y \cosh G\, (dt+V_{\varphi} d\varphi)^2+\frac{1}{2y \cosh G} (dr^2 +dy^2 +r^2 d\varphi^2) + y e^{G} d\Omega_{3}^2+ y e^{-G} d\tilde{\Omega}_{3}^2
\eeq 
where the function $G$ and $V_{\varphi}$ depend only on the radial coordinate $r$ on the plane (due to rotational invariance) and $y$. They take the following form
\beq
\begin{aligned}
 \frac{1}{2}\tanh G(r,y) \equiv z (r,y)  &= \frac{1}{2}+\sum_{k}^{p}(-1)^{k+1}\left[\frac{ r^{2}-r_{k}^2+y^2}{2 \sqrt{(r^2+r_{i}^2+y^2)^2 -4 r^2 r_{k}^2}} -\frac{1}{2} \right]\\
 V_{\varphi} (r,y) &=\sum_{k}^{p}(-1)^{k}\left[\frac{ r^{2}+r_{k}^2+y^2}{2 \sqrt{(r^2+r_{i}^2+y^2)^2 -4 r^2 r_{k}^2}} -\frac{1}{2} \right]\,,
\end{aligned}
\eeq
with $r_{k}>r_{k+1}$.
For $y=0$, $z(r,0)$ indeed can only take two values $\pm \frac{1}{2}$ and therefore this plane is characterized by a set of $p$ concentric rings with colors alternating in black and white depending on the value of $z$. The simplest of these configurations is a single black disk ($p=1$) and that corresponds to the $AdS_{5}\times S^{5}$ solution (see figure \ref{fig:LLM}). The radius of such disk is identified with the radius of AdS (more precisely, $r_{1}=R_{\text{AdS}}^2$).
The next to simplest case is obtained by setting $p=2$ with the resulting picture being a white annulus on a black background, see figure \ref{fig:LLM}. 

The LLM backgrounds can be understood as global coordinates analogs of Coulomb branch solutions (the near horizon geometries of  stacks of parallel D3 branes in asymptotically flat space), to which  they indeed reduce in a suitable
 limit~\cite{Lin:2004nb}.  As we move from the boundary to the interior along the radial direction, the effective number of D3 branes (the effective rank of the gauge group) decreases. From the viewpoint of the theory living on the boundary of the full space, this is a standard RG flow. A comprehensive analysis of the standard holographic dictionary for the LLM solutions was presented in~\cite{Skenderis:2007yb}.
 But we can also imagine describing the flow from the viewpoint of the ``interior'' effective field theory with the smaller number of branes. On general grounds, 
the low-energy EFT will be of the form described in this paper, i.e.~SYM theory perturbed by the leading irrelevant deformation ${\cal O}_8$.

Focussing on the simplest non-trivial case $p=2$, we can form a dimensionless parameter by considering the ratio of the two radii of the outer and inner circle, $ r_{2}^2/r_{1}^2$. As this parameter goes to zero, we recover the $AdS_{5}\times S^{5}$ solution in analogy to the IR limit of the irrelevant deformation. It is then tempting to consider this geometry as a potential candidate for an holographic dual to the irrelevant deformation and identify the ratio of the radii with the dimensionless combination of the coupling constant and sphere radius $\ell$ as follows
\beq
\frac{r_{2}^2}{r_{1}^2} \sim \frac{h}{\ell^{4}}\,.
\eeq
The LLM geometries are however asymptotically AdS and  we cannot recover the flat space asymptotics similar to the D3-brane background. This is a consequence of the deformed theory being defined on $S^3 \times \mathbb{R}$ rather than in flat space.

The classical integrability properties of the LLM solutions have been investigated in \cite{Chervonyi:2013eja} for the point-like limit of strings in this background\footnote{See also \cite{deMelloKoch:2018tlb} for some tests of integrability on the LLM backgrounds based on the underlying central extended algebra.}. The separability of the equations for massless geodesics  was studied and the outcome shows that only the $AdS_{5}\times S^{5}$ background (and its pp-wave limit) can be integrable. 
Nevertheless, we should remark that our perturbative analysis was performed for operators with very large R charge, which is expected to be related to very long strings, far from the point-like limit. In case  this LLM solution is in fact dual to the irrelevant deformation studied here, one logical possibility to make this result compatible with our findings could be that finite size effects might lead to the integrability breakdown. An alternative but related scenario, is that one might hope to find at least some perturbatively closed subsectors that remain integrable while the full theory is not. We hope to further investigate the relation between LLM and our deformation.

\section{Discussion}

\label{conclusions_sec}
In this paper we have studied the leading irrelevant deformation of $\mathcal{N}=4$ SYM preserving half of its 32 supercharges on $S^3 \times \mathbb{R}$. Thanks to an off-shell formalism well suited for the symmetries of our problem,  we  have managed to write down the full deformed classical action  in  closed form. At the quantum level and in the planar limit, we have studied in conformal perturbation theory the generator of time translations along the cylinder axis and its integrability properties. For the $\mathfrak{su}(2|2)\ltimes \mathbb{R}$ closed subsector of the theory, purely algebraic considerations are enough to fully constraint the form of the supersymmetric generators up to two loops. As an outcome, we have found an enhancement of the symmetries so that the time-translation generator reduces to the spin chain Hamiltonian of $\mathcal{N}=4$ SYM. Therefore this does not settle the question of integrability of the theory, which remains open.

Since the method employed here to probe integrability only involved an extensive use of the underlying symmetries, one can formulate the more general  question of finding long-range integrable spin chains with $\mathfrak{psu}(2|2)^2 \ltimes \mathbb{R}^2$ symmetry. At higher orders this is a highly nontrivial problem and a generalization of the tools developed in \cite{Bargheer:2009xy} for bosonic spin chains might be a possible route. In particular, two significant differences in our setup compared to the examples studied there are the dynamic nature of the spin chains (by this we mean that the spin chain length fluctuates) and the fact that the generators of the algebra are themselves corrected at quantum level. In particular, the latter  makes the global symmetry not manifest at the level of the Hamiltonian.

On the string theory side of the problem, we have suggested a potential candidate for a holographic dual of the deformed theory, motivated by the match of symmetries and a common IR limit: the bubbling geometries of Lin, Lunin and Maldacena.  To sharpen this suggestion, it would be interesting to study the first perturbation in the small parameter ${r_{2}^2}/{r_{1}^2} $ of the LLM background around  $AdS_{5}\times S^{5}$ and relate it to the field theory leading irrelevant operator. More generally, numerous  integrable deformations of the $AdS_{5}\times S^{5}$ sigma model have been thoroughly studied  (see for instance \cite{Orlando:2019his} for a review). A natural question is whether one can find among those some further examples which have the same symmetries studied here and are also type IIB solutions.\footnote{Recently, a new class of integrable deformations of sigma-models has been proposed in \cite{Sfondrini:2019smd} for which the resulting world-sheet S-matrix gets modified only by a CDD factor in line with the analysis of section \ref{longrange}. However, when applied to the $AdS_5 \times S^{5}$ example, this modification of the S-matrix seems incompatible with our perturbative results.}

Finally, we would like to mention a natural generalization of our setup, inspired by the analysis of~\cite{Skenderis:2007yb}.\footnote{We thank Kostas Skenderis for  asking about this generalization.}
The asymptotically flat D3-brane solution (\ref{D3metric}) can be generalized to any harmonic function $H (x_I)$. Terms that fall-off as $r \to \infty$ are associated with the Coulomb branch of  ${\cal N}=4$ SYM. The constant term in $H$ is associated with  ${\cal O}_8$, and  the terms that go as $1/ r^k$ with its Kaluza-Klein cousins,  of the schematic form
${\cal O}_{8+k} \sim {\rm Tr} F^4 X^k$.  
This more general setup in  flat space has a natural counterpart on $S^3 \times \mathbb{R}$, where we  deform ${\cal N}=4$ SYM by a general sum of  the irrelevant
one-half BPS operators ${\cal O}_{8+k}$. In order to preserve 16 supercharges on $S^3 \times \mathbb{R}$, we need to align the operators in R-symmetry space so that they are charged under the same  $U(1)_{\mathcal J} \subset SO(6)_R$, which would then be broken by adding them to the action. The classical action is an immediate generalization of what we wrote above, with the additional terms corresponding to 
$\int d^8\theta Tr \bar Z^{4 + k}$ in our superspace language. It would be interesting to repeat the spin chain analysis in this more general case. As the preserved symmetry is smaller, it is conceivable that one one might find a genuine deviation from the ${\cal N}=4$ result already at two loops.

\section*{Acknowledgments}
The authors would like thank Niklas Beisert, Nikolay Bobev, Fran\c{c}ois Delduc,  Fridrik Frey Gautason, Shota Komatsu, Marc Magro, Juan Maldacena, Carlo Meneghelli, M\'{a}rk Mezei,  Alessandro Sfondrini, Kostas Skenderis and Yifan Wang for discussions.
The work of L.R.  is supported in part by NSF grant \#~PHY-1915093.

\appendix

\section{Notations and conventions}\label{notations}
We realize $\mathcal N=4$ super-Yang-Mills theory in the usual fashion as the trivial dimensional reduction of ten-dimensional $\mathcal N=1$ SYM on $\mathbb T^6$. Correspondingly, the ten-dimensional vielbein indices are subjected to a $4+6$ split. Additionally, we will separate the latter six indices into a group of four and a group of two. To keep track of the various ranges the indices take values in, we use the following notations:
\begin{eqnarray*}
M,N,\ldots &=&0,1,2,3,4,5,6,7,8,9 \\
m,n,\ldots &=&0,1,2,3 \\
\hat m,\hat n,\ldots &=&1,2,3 \\
I,J,\ldots &=&4,5,6,7,8,9\\
i,j,\ldots &=&4,5,6,7\\
u,v\ldots &=&8,9
\end{eqnarray*}
The curved-space indices associated with $m,n,\ldots$ and $\hat m, \hat n, \ldots$ are denoted by their corresponding Greek letters $\mu,\nu,\ldots$ and $\hat\mu,\hat\nu,\ldots$

The vielbein indices $M,N,\ldots$ are acted on by $SO(1,9)$ local Lorentz rotations. This symmetry groups breaks into $SO(1,3)\times SO(6)_R$ upon performing the $4+6$ split. Here $SO(1,3)$ are the standard four-dimensional local Lorentz rotations, while $SO(6)_R$ is the $\mathcal N=4$ R-symmetry. The latter is broken further to $SO(4)\times SO(2)$ by the $4+2$-split. We denote the generator of $\mathfrak{so}(2)\cong \mathfrak{u}(1)$ as $\mathcal J$.

We represent the ten-dimensional gamma-matrices in a guise most natural from the point of view of the $SO(1,3)\times SO(4)\times SO(2) \subset SO(1,9)$ subgroup. Concretely,
\begin{align}\label{10dGamma}
\Gamma_m &= \unit_2\otimes \unit_4 \otimes \gamma_m\\
\Gamma_i &= \unit_2\otimes \tilde\gamma_{i-3} \otimes \gamma_5\\
\Gamma_u &=\tau_{u-7} \otimes \gamma_5 \otimes \gamma_5\;.
\end{align}
Here $\unit_d$ is the $d$-dimensional unit-matrix, and we used the standard four-dimensional $\gamma$-matrices in $(1,3)$-signature, and denoted those in $(0,4)$ signature as $\tilde \gamma$. Using the standard Pauli-matrices $\tau_{1}=\left(\begin{smallmatrix} 0 & 1\\ 1& 0 \end{smallmatrix}\right)$, $\tau_{2}=\left(\begin{smallmatrix} 0 & -i\\ i& 0 \end{smallmatrix}\right)$, and  $\tau_{3}=\left(\begin{smallmatrix} 1 & 0\\ 0& -1 \end{smallmatrix}\right)$, we write them in terms of the $\sigma$-matrices
\begin{align}
&\sigma_0=-i \unit_2\;, \qquad \sigma_1=-i \tau_1\;, \qquad \sigma_2=-i \tau_2\;, \qquad \sigma_3=-i \tau_3\;, \qquad \sigma_4=\unit_2\;,\\
&\bar\sigma_0=-i \unit_2\;, \qquad \bar\sigma_1=+i \tau_1\;, \qquad \bar\sigma_2=+i \tau_2\;, \qquad \bar\sigma_3=+i \tau_3\;, \qquad \bar\sigma_4=\unit_2\;,
\end{align}
as
\begin{equation}
\gamma_m = \begin{pmatrix}
0 & \sigma_m \\
\bar\sigma_m & 0
\end{pmatrix}\;, \qquad \text{and}\qquad \tilde\gamma_{i-3} = \begin{pmatrix}
0 & \sigma_{i-3} \\
\bar\sigma_{i-3} & 0
\end{pmatrix}\;.
\end{equation}
Finally, $\gamma_5 = -i \gamma_0\gamma_1\gamma_2\gamma_3 = \tilde \gamma_{1}\tilde\gamma_2\tilde\gamma_3\tilde\gamma_4$ is the standard four-dimensional chirality matrix. We endow the $\sigma$-matrices with spinor-indices, each taking two values, in the standard manner:
\begin{equation}
(\sigma_m)_{\mathsf{a}\dot{\mathsf{a}}}\;, \qquad (\bar\sigma_m)^{\dot{\mathsf{a}}\mathsf{a}}\;, \qquad \text{and}\qquad (\sigma_{i-3})_{a\dot{a}}\;, \qquad (\bar\sigma_{i-3})^{\dot{a}a}\;.
\end{equation}
As usual, these can be used to trade vector indices with spinorial indices. Here we notationally distinguished the $SU(2)_a\times SU(2)_{\dot a}\simeq SO(4)_R $ R-symmetry indices $a, \dot a$, from the spatial spinor indices $\mathsf{a}, \dot{\mathsf{a}}$. To avoid clutter, we often omit the subtractions of three (seven) units of the indices $i,j,\ldots$ ($u,v,\ldots$) that keep their values in their standard range. 

The ten-dimensional chirality matrix is given by
\begin{equation}\label{chiralityMatrix}
\Gamma_{11} = \Gamma_0\Gamma_1\Gamma_2\Gamma_3\Gamma_4\Gamma_5\Gamma_6\Gamma_7\Gamma_8\Gamma_9= - \tau_3\otimes\gamma_5\otimes \gamma_5\;.
\end{equation}
We will be interested in Weyl spinors. The ten-dimensional charge conjugation matrix reads
\begin{equation}
C_{10d} = i \tau_2\otimes C_{4d}\otimes C_{4d}\;, \qquad C_{10d}\, \Gamma_M\, C_{10d}^{-1} = -\Gamma_M^\trans\;, \qquad C_{10d}^\trans = -C_{10d}\;,
\end{equation}
where
\begin{equation}
C_{4d} = i \gamma_0\gamma_2 = \tilde \gamma_4\tilde\gamma_2\;, \qquad C_{4d}\, \gamma_m\, C_{4d}^{-1} = -\gamma_m^\trans\;, \qquad C_{4d}\, \tilde \gamma_i\, C_{4d}^{-1} = -\tilde \gamma_i^\trans\;, \qquad C_{4d}^\trans = -C_{4d}\;.
\end{equation}
The spinors we encounter satisfy the Majorana reality condition:
\begin{equation}\label{Majcond}
\bar\Psi = \Psi^\dagger \, i \Gamma_0 = \Psi^\trans\, C_{10d}\;.
\end{equation}

The 32-component index of a Dirac spinor $\Psi$ can be decomposed similarly to \eqref{10dGamma}
\begin{equation}\label{32compspinor}
\Psi\left[\mathfrak s \otimes \begin{pmatrix}
(l)a\\
(u)\dot a
\end{pmatrix}\otimes\begin{pmatrix}
(l)\mathsf{a}\\
(u)\dot{\mathsf{a}}
\end{pmatrix}\right]\;,
\end{equation}
where $\mathfrak s = \pm$ and $(l),(u)$ indicate a lower or upper position of the index. Note that the $U(1)_\mathcal{J}$ generator acts on spinors as
\begin{equation}\label{U1Jactiononspinors}
\mathcal J = -\frac{i}{2} \Gamma_{89} = \frac{1}{2} \tau_3 \otimes \unit_4 \otimes \unit_4\;,
\end{equation}
hence the sign $\mathfrak s = \pm$ is (proportional to) the $U(1)_\mathcal{J}$ charge. In view of the tensor product structure of the chirality matrix in \eqref{chiralityMatrix}, the Weyl condition demands the product of the chirality of the three factors be of definite sign. Hence we have
\begin{align}\label{componentsWeylspinor}
&\text{components of positive chirality spinor:} &&\psi_{+a}^{\dot{\mathsf{a}}}\;, \qquad \psi_{+\mathsf{a}}^{\dot{a}}\;, \qquad \psi_{-a\mathsf{a}}\;, \qquad \psi_{-}^{\dot{a}\dot{\mathsf{a}}}\;,\\
&\text{components of negative chirality spinor:}  &&\psi_{-a}^{\dot{\mathsf{a}}}\;, \qquad \psi_{-\mathsf{a}}^{\dot{a}}\;, \qquad \psi_{+a\mathsf{a}}\;, \qquad \psi_{+}^{\dot{a}\dot{\mathsf{a}}}\;.
\end{align}

The reality property \eqref{Majcond} implies that
\begin{equation}\label{Majcondincomp}
(\psi_{\mathfrak s a\mathsf{a}})^* = - \epsilon^{\mathfrak s \mathfrak t}\, \epsilon^{ab}\, \epsilon_{\dot{\mathsf{a}}\dot{\mathsf{b}}} \, \psi_{\mathfrak t b}^{\dot{\mathsf{b}}}\;, \qquad (\psi_{\mathfrak s \mathsf{a}}^{\dot a})^* = - \epsilon^{\mathfrak s \mathfrak t}\, \epsilon_{\dot a\dot b}\, \epsilon_{\dot{\mathsf{a}}\dot{\mathsf{b}}} \, \psi_{\mathfrak t}^{\dot b\dot{\mathsf{b}}}\;.
\end{equation}
Here $\dot {\mathsf a}$ on the right-hand side takes the same numerical value as $\mathsf a$ on the left-hand side. Furthermore, the $\epsilon$-tensors are
\begin{equation}
\epsilon_{ab} = \begin{pmatrix}
0 & -1 \\
1 & 0
\end{pmatrix}\;, \qquad \epsilon^{ab} = \begin{pmatrix}
0 & 1 \\
-1 & 0
\end{pmatrix}\;,
\end{equation}
and indices are raised and lowered as
\begin{equation}
v_a = \epsilon_{ab}v^b \;, \qquad v^a = \epsilon^{ab} v_b\;,
\end{equation}
and similarly with dotted indices, and for the other kinds of indices.

\section{\texorpdfstring{$\mathcal N=4$}{N=4} SYM on \texorpdfstring{$S^3\times \mathbb R$}{S³xR}}\label{app:N=4onS3S1}
Euclidean $S^3\times \mathbb R$ is conformally flat, hence, via a Weyl transformation, it is easy to convince oneself that it supports all 16+16 $\mathcal N=4$ super(conformal) charges, which are the fermionic generators of the superalgebra $\mathfrak{psu}(4^*|4^*)$. Next, we can of course Wick rotate along $\mathbb R$ without changing the number of preserved supercharges. In other words, we can find 16+16 independent solutions to the conformal Killing spinor equation
\begin{equation}\label{KSE}
D_\mu \varepsilon = \Gamma_\mu \tilde \varepsilon\;,
\end{equation}
for $\tilde \varepsilon = \frac{1}{4}\Gamma^\mu D_\mu \varepsilon$. Here $\varepsilon$ is a Majorana-Weyl spinor, which we choose to be of positive chirality. Furthermore, 
\begin{equation}
D_\mu \varepsilon = \partial_\mu \varepsilon + \frac{1}{4} \omega_\mu^{mn} \Gamma_{mn} \varepsilon\;,
\end{equation} 
where $\omega_{\mu}^{mn}$ are components of the spin-connection.

Let us describe these solutions explicitly. We take the metric on $S^3\times \mathbb R$ to be
\begin{equation}
ds^2 =  - dt^2 + \ell^2 \big(\cos^2\vartheta d\varphi^2 + \sin^2\vartheta d\chi^2 + d\vartheta^2\big)\;.
\end{equation}
The coordinates $\varphi,\chi$ are periodic with period $2\pi$ and $\vartheta$ runs over $[0,\pi/2]$. We choose the vielbein
\begin{equation}\label{vielbein}
e^0=dt\;,\qquad e^1=\ell\cos\vartheta d\varphi\;, \qquad e^2=\ell \sin\vartheta d\chi\;, \qquad e^3=\ell d\vartheta\;, 
\end{equation}
and compute
\begin{equation}
e_0 = \partial_t \;, \qquad e_1=\frac{1}{\ell\cos\vartheta}\partial_\varphi\;, \qquad e_2=\frac{1}{\ell\sin\vartheta}\partial_\chi\;, \qquad e_3 = \frac{1}{\ell}\partial_\vartheta\;.
\end{equation}
The nonzero components of the spin-connection are
\begin{equation}\label{spinconnection}
\omega^{13}=-\omega^{31}=-\sin\theta d\varphi\;, \qquad \omega^{23}=-\omega^{32}=\cos\theta d\chi\;.
\end{equation}

As mentioned above, a convenient way to find the solutions to the Killing spinor equation \eqref{KSE} is to consider the Euclidean flat space solutions, i.e., $\varepsilon = (\hat\varepsilon_s + x^\mu \Gamma_\mu \hat\varepsilon_c)$, with constant $\hat\epsilon_s,\hat\epsilon_c$ satisfying the appropriate reality and chirality conditions, then to perform the Weyl map to $S^3\times \mathbb R$, and finally apply a frame rotation to align the frame with the one defined in \eqref{vielbein}. Alternatively, we can straightforwardly solve the equation directly on $S^3\times \mathbb R$, following, for example, \cite{Peelaers:2014ima}. Either way, the solution can be most conveniently written in terms of the two-component spinors
\begin{equation}\label{kappaspinors}
\kappa_{s\tilde s} = \frac{1}{2}\begin{pmatrix} 
e^{\frac{i}{2}(s\chi + \tilde s \varphi -s \tilde s \vartheta)}\\
-s e^{\frac{i}{2}(s \chi + \tilde s \varphi + s\tilde s \vartheta)}
\end{pmatrix}\;, \qquad \text{for}\qquad s,\tilde s = \pm 1\;,
\end{equation}
which satisfy
\begin{equation}
\nabla_{\hat\mu} \kappa_{s\tilde s} = -\frac{i s\tilde s}{2\ell} \tau_{\hat \mu}\kappa_{s\tilde s}\;.
\end{equation}

The components of the most general spinor $\varepsilon$ solving the Killing spinor equation \eqref{KSE} are
\begin{align}
&\varepsilon_{+a}^{\dot{\mathsf{a}}} = \sum_{s_a,{\tilde s}_a=\pm1} c_a^{(s_a,{\tilde s}_a)}\  e^{\frac{i s_a {\tilde s}_a t}{2\ell}} (\kappa_{s_a{\tilde s}_a})^{\dot{\mathsf{a}}}\;, \qquad && \varepsilon_{+\mathsf{a}}^{\dot{a}} = \sum_{s_{\dot a},{\tilde s}_{\dot a}=\pm1} c^{\dot{a}}_{(s_{\dot a},{\tilde s}_{\dot a})}\  e^{-\frac{i s_{\dot a} {\tilde s}_{\dot a} t}{2\ell}} (\kappa_{s_{\dot a}{\tilde s}_{\dot a}})_{{\mathsf{a}}}\\
&\varepsilon_{-a\mathsf{a}} = \sum_{s_a,{\tilde s}_a=\pm1} d_a^{(s_a,{\tilde s}_a)}\  e^{-\frac{i s_a {\tilde s}_a t}{2\ell}} (\kappa_{s_a{\tilde s}_a})_{{\mathsf{a}}}\;, \qquad && \varepsilon_{-}^{\dot{a}\dot{\mathsf{a}}} = \sum_{s_{\dot a},{\tilde s}_{\dot a}=\pm1} d^{\dot{a}}_{(s_{\dot a},{\tilde s}_{\dot a})}\  e^{\frac{i s_{\dot a} {\tilde s}_{\dot a} t}{2\ell}} (\kappa_{s_{\dot a}{\tilde s}_{\dot a}})^{\dot{\mathsf{a}}}\;,
\end{align}
where the coefficients $c_a^{(s_a,{\tilde s}_a)}$, $d_a^{(s_a,{\tilde s}_a)}$, $c^{\dot{a}}_{(s_{\dot a},{\tilde s}_{\dot a})}$ and $d^{\dot{a}}_{(s_{\dot a},{\tilde s}_{\dot a})}$  are complex constants satisfying reality properties implementing the Majorana constraints in \eqref{Majcondincomp}:
\begin{equation}
(d_a^{(s_a,{\tilde s}_a)})^* = -s_a\ \epsilon^{ab}\, c_b^{(-s_a,-{\tilde s}_a)}\;, \qquad  (c^{\dot{a}}_{(s_{\dot a},{\tilde s}_{\dot a})})^* =  s_{\dot a}\ \epsilon_{\dot a \dot b}\, d^{\dot{b}}_{(-s_{\dot a},-{\tilde s}_{\dot a})}\;.
\end{equation}
In total, we recover the expected 32 (real) supercharges. 

The on-shell transformation rules of the $\mathcal N=4$ vector multiplet can be easily written down (see, \textit{e.g.}, \cite{Pestun:2007rz})
\begin{align}\label{onshelltransformation1}
\delta A_M &= \bar \Psi \Gamma_M \varepsilon\;,\\\label{onshelltransformation2}
\delta \Psi &= -\frac{1}{2} F_{MN} \Gamma^{MN}\varepsilon - \frac{1}{2} \Phi_I \Gamma^{\mu I} D_\mu \varepsilon\;.
\end{align}
Here $A_M=(A_\mu,\Phi_I)$ contains the gauge field $A_\mu$ and the standard six scalar fields $\Phi_I$, and $\Psi$ is a Majorana-Weyl fermion of positive chirality, as $\varepsilon$, and thus with components as in \eqref{componentsWeylspinor}. All are valued in the adjoint representation of the gauge group, and we take the generators to be antihermitian. The spinor $\varepsilon$ is chosen Grassmann-even, so that $\delta$ is a linear combination of (Grassmann-odd) supercharges, and it satisfies the Killing spinor equation \eqref{KSE}. One can verify that
\begin{equation}
\delta^2 = \mathcal L^A_{\xi} + \text{Scale}(w) + R_{SO(6)_R}(\Theta_{IJ}) + \text{eom}[\Psi]\;,
\end{equation}
where $\mathcal L^A_{\xi}$ denotes the gauge covariant Lie derivative along the conformal Killing vector field $\xi^M = \bar\varepsilon \Gamma^M\varepsilon$, $\text{Scale}(w)$ is a local scale transformation with parameter $w = \bar\varepsilon\tilde\varepsilon$, and finally $R_{SO(6)}(\Theta_{IJ})$ represents an $SO(6)_R$ R-symmetry rotation with parameter $\Theta_{IJ} = 2 \bar\varepsilon\Gamma_{IJ}\tilde\varepsilon $.

The Lagrangian invariant (up to total derivative terms) under the on-shell transformation rules \eqref{onshelltransformation1}-\eqref{onshelltransformation2} reads
\begin{equation}
L_{\text{YM}}^{(\text{on-shell})} = \frac{1}{2} F_{MN}F^{MN} + \bar \Psi \Gamma^M D_M \Psi + \frac{R}{6} \Phi_I \Phi^I\;,
\end{equation}
where $R$ is the scalar curvature of space-time, which for $S^3\times \mathbb R$ is $R= \frac{6}{\ell^2}$.

\section{Rigid subalgebra}\label{app:rigidsub}
Let us consider in some more detail the Killing spinors describing the supercharges of the ``rigid'' subalgebra presented in \eqref{superisometries}, {i.e.},
\begin{equation}\label{superisometriesapp}
\mathfrak{psu}(2|2) \times \mathfrak{psu}(2|2) \ltimes \mathbb{R}^2\;,
\end{equation}
which are selected by the requirement that they commute with $\mathcal H-\mathcal J$. First, we remark that the spinors $\kappa_{s\tilde s}$ introduced in equation \eqref{kappaspinors} can be organized in doublets of $SU(2)_{\mathcal L} \times SU(2)_{\mathcal R}$ as follows: $(\kappa_{+-},\kappa_{-+})$ is a doublet of $SU(2)_{\mathcal L}$, and similarly $(\kappa_{--},\kappa_{++})$ is a doublet of $SU(2)_{\mathcal R}$. The $SU(2)_{\mathcal L}$ and $SU(2)_{\mathcal R}$ Killing vectors $\mathcal L_A,\mathcal R_A$ can be constructed as the usual Killing spinor bilinears. One finds
\begin{align}
&\mathcal L_{3} = (\bar\kappa_{+-}\ \tau_{\hat m}\  \kappa_{-+})\, \ell e_{\hat m} = \frac{i}{2}(\partial_\varphi - \partial_\chi)\;,\\
&\mathcal L_{+} = (\bar\kappa_{+-}\ \tau_{\hat m}\  \kappa_{+-})\, \ell e_{\hat m} = \frac{1}{2}e^{-i(\varphi-\chi)}(-\tan\vartheta\ \partial_\varphi - \cot\vartheta\ \partial_\chi+i \partial_\vartheta)\;,\\
&\mathcal L_{-} = (\bar\kappa_{+-}\ \tau_{\hat m}\  \kappa_{+-})\, \ell e_{\hat m} = \frac{1}{2}e^{i(\varphi-\chi)}(-\tan\vartheta\ \partial_\varphi - \cot\vartheta\ \partial_\chi-i \partial_\vartheta)\;,
\end{align}
and
\begin{align}
&\mathcal R_{3} = (\bar\kappa_{++}\ \tau_{\hat m}\  \kappa_{--})\, \ell e_{\hat m} = \frac{i}{2}(\partial_\varphi + \partial_\chi)\;,\\
&\mathcal R_{+} = (\bar\kappa_{--}\ \tau_{\hat m}\  \kappa_{--})\, \ell e_{\hat m} = \frac{1}{2}e^{-i(\varphi+\chi)}(\tan\vartheta\ \partial_\varphi - \cot\vartheta\ \partial_\chi-i \partial_\vartheta)\;,\\
&\mathcal R_{-} = (\bar\kappa_{++}\ \tau_{\hat m}\  \kappa_{++})\, \ell e_{\hat m} = \frac{1}{2}e^{i(\varphi+\chi)}(\tan\vartheta\ \partial_\varphi - \cot\vartheta\ \partial_\chi+i \partial_\vartheta)\;,
\end{align}
where $\bar \kappa = (\tau_2\, \kappa)^\trans$. It is standard to define the left-invariant frame $e_{(\mathcal L)}^A$ on the three-sphere as the frame whose dual framevectors are the vectorfields $\mathcal L_A$. This frame, completed to a frame on $S^3\times \mathbb R$ in the obvious manner, can be obtained from our frame \eqref{vielbein} by a local Lorentz rotation $M_{(\mathcal L)}$: $e_{(\mathcal L)}^A = (M_{(\mathcal L)})^A_m e^m$. We denote its corresponding rotation on chiral spinors as $m_{(\mathcal L)}$ and on anti-chiral spinors as $\tilde m_{(\mathcal L)}$.\footnote{Since the local Lorentz rotation is an element of $SO(3)\subset SO(3,1)$, the same matrix acts on chiral and anti-chiral spinors: $m_{(\mathcal L)}=\tilde m_{(\mathcal L)}$ as matrices.} In particular, one finds that in the new frame the $SU(2)_{\mathcal L}$ doublet of spinors $(\kappa_{+-},\kappa_{-+})$ is constant and diagonal in the spinor index and $SU(2)_{\mathcal L}$ doublet index. In other words,
\begin{equation}
m_{(\mathcal L)} \cdot \kappa_{+-} = \begin{pmatrix}
c\\ 0
\end{pmatrix} \;, \qquad m_{(\mathcal L)}\cdot \kappa_{-+} = \begin{pmatrix}
0\\ c'
\end{pmatrix}\;,
\end{equation}
for some constants $c,c'$. Similar statements can be made about the right-invariant frame and the $SU(2)_{\mathcal R}$ doublet of spinors $(\kappa_{--},\kappa_{++})$.

With these preparations, we come to the description of the Killing spinors associated with the raising supercharges of the rigid subalgebra; the lowering supercharges are described by complex conjugation. The associated Killing spinors are described by the eigenvalues $\mathcal H=\mathcal J = -\frac{1}{2}$. Recalling from \eqref{U1Jactiononspinors} the action of $U(1)_\mathcal{J}$ on spinors and setting $\mathcal H = i\ell \partial_t$, we are thus focusing on the Killing spinors
\begin{equation}
\varepsilon_{-a\mathsf{a}} =  e^{\frac{i t}{2\ell}} (d_a^{(+-)}\  (\kappa_{+-})_{{\mathsf{a}}}+d_a^{(-+)}\  (\kappa_{-+})_{{\mathsf{a}}})\;, \qquad  \varepsilon_{-}^{\dot{a}\dot{\mathsf{a}}} =  e^{\frac{i t}{2\ell}} ( d^{\dot{a}}_{(++)}\   (\kappa_{++})^{\dot{\mathsf{a}}}+d^{\dot{a}}_{(--)}\   (\kappa_{--})^{\dot{\mathsf{a}}})\;.
\end{equation}
Note that the former is a linear combination of the components of the $SU(2)_{\mathcal L}$ doublet of spinors, while the latter of the $SU(2)_{\mathcal R}$ doublet of spinors. To emphasize these transformation properties, we introduce the index $\alpha$ ($\dot\alpha$) for $SU(2)_{\mathcal L}$ ($SU(2)_{\mathcal R}$) and write for the doublet of terms
\begin{equation}
\varepsilon_{-a\mathsf{a}\alpha}\;, \qquad \varepsilon_{-}^{\dot{a}\dot{\mathsf{a}}\dot{\alpha}}\;.
\end{equation}

Per the previous discussion, in the left-invariant frame, it is redundant to keep both the index $\mathsf a$ and $\alpha$ on $\varepsilon_{-a\mathsf{a}\alpha}$ as they can be identified, and similarly for the indices $\dot{\mathsf{a}},\dot{\alpha}$ of the other Killing spinor in the right-invariant frame. While our frame is neither left- nor right-invariant, we will use notations adopted to these respective frames and simply write
\begin{equation}
\varepsilon_{-a\alpha}\;, \qquad \varepsilon_{-}^{\dot{a}\dot{\alpha}}\;,
\end{equation}
This notation can thus be read as either having suppressed the explicit spinor indices, or, more usefully and as we will use it, as having kept implicit the frame-rotations $m_{(\mathcal L)}$ and $\tilde m_{(\mathcal R)}$.

\section{Off-shell realization}\label{app:off-shell}

It was shown in \cite{Berkovits:1993hx,Evans:1994np} that one can simultaneously close off-shell at most nine supersymmetries of $\mathcal N=4$ super Yang-Mills.\footnote{More generally, one can preserve nine supersymmetries and nine special conformal supersymmetries.} In general, the off-shell theory preserves only a subgroup of the spatial and $R$-symmetry group of the original theory. Following \cite{Berkovits:1993hx,Evans:1994np}, we would like to realize off-shell and in a linear manner the eight raising supercharges of the rigid subalgebra in \eqref{superisometries}, while preserving the full rigid bosonic subalgebra.

The first step is to introduce seven auxiliary fields $K^\ell$ and modify the supersymmetry transformation rules \eqref{onshelltransformation1}-\eqref{onshelltransformation2} to
\begin{align}\label{offshelltransformation1}
\delta A_M &= \bar \Psi \Gamma_M \varepsilon\;,\\\label{offshelltransformation2}
\delta \Psi &= -\frac{1}{2} F_{MN} \Gamma^{MN}\varepsilon - \frac{1}{2} \Phi_I \Gamma^{\mu I} D_\mu \epsilon - K^\ell \nu_\ell\;,\\\label{offshelltransformation3}
\delta K_{\ell} &= - \bar\nu_\ell \Gamma^M D_M \Psi\;.
\end{align}
Here $\nu_\ell$ are seven auxiliary spinors. To ensure that the algebra closes off-shell they must satisfy
\begin{equation}\label{constraintsnu}
\bar\varepsilon \Gamma^M \nu_\ell = 0\;, \qquad \bar\nu_\ell \Gamma^M \nu_{\ell'} = \delta_{\ell\ell'}\ \bar\varepsilon \Gamma^M \varepsilon\;,
\end{equation}
for each Killing spinor $\varepsilon$ we would like to preserve off-shell. Using that the Killing spinors we have selected satisfy $\mathcal J = -\frac{1}{2}$, {i.e.},
\begin{equation}
i \Gamma_{89} \varepsilon = \varepsilon\;,
\end{equation}
one can easily verify that a solution to the constraints \eqref{constraintsnu} is given by
\begin{equation}
\nu_{\hat m} = i\Gamma_{\hat{m}0} \varepsilon\;, \qquad \text{and} \qquad \nu_{j} = i\Gamma_{j0} \varepsilon\;.
\end{equation}
Recall from appendix \ref{notations} that $\hat m = 1,2,3$ and $j=4,5,6,7$. Remarkably, this $7 = 3 + 4$ split preserves the full isometries of $S^3 \times \mathbb R$ and the full $SO(4)_R \times  U(1)_\mathcal{J}$ $R$-symmetry, if we declare that the corresponding auxiliary fields $K^{\hat{\mu}}$ transform as a spatial vector on $S^3$ and $K^i$ as a vector of $SO(4)_R$.

It is useful to write the supersymmetry variations of these eight (raising) supercharges of the rigid subalgebra in detail. To do so, we first observe that they satisfy
\begin{equation}\label{KSequationplus}
D_\mu \epsilon = -\frac{i}{2\ell}\Gamma_\mu \Gamma_0 \epsilon\;.
\end{equation}
Then we find
\begin{align}\label{offshelltransformation1bis}
\delta A_M &= \bar \Psi \Gamma_M \varepsilon\;,\\\label{offshelltransformation2bis}
\delta \Psi &= -\frac{1}{2} F_{MN} \Gamma^{MN}\varepsilon - \frac{i}{\ell} \Phi^I \Gamma_{I0}\varepsilon - i K^{\hat m} \Gamma_{\hat{m}0} \varepsilon - i K^{j} \Gamma_{j0} \varepsilon\;,\\\label{offshelltransformation3bis}
\delta K_{\hat m} &= -i  D_M \bar\Psi \Gamma^M \Gamma_{\hat m 0}\varepsilon \;.\\
\delta K_{j} &= -i  D_M \bar\Psi \Gamma^M \Gamma_{j 0}\varepsilon \;.
\end{align}
Note that the variation of $\Psi$ has an explicit $1/\ell$ correction. We should also keep in mind that these variations are only valid for the eight supercharges we selected.

\section{\texorpdfstring{$\bar Z$}{Zbar} superfield}\label{app:Zbarsuperfield}
The complete $\bar Z$-superfield is given by
\begin{scriptsize}
\begin{align*}
&\bar Z(\theta_{a\alpha}, \tilde\theta^{\dot a\dot\alpha})= \bar Z - 2i \epsilon^{ab}\epsilon^{\alpha\beta}\Psi_{-a\alpha} \, \theta_{b\beta} - 2i \epsilon_{\dot a \dot b}\epsilon_{\dot\alpha\dot \beta} \Psi_{-}^{\dot a\dot\alpha} \, \tilde\theta^{\dot b\dot\beta} \displaybreak[0]\\
&\quad+i \epsilon^{ab}\epsilon^{\alpha\beta}\left(\frac{1}{2}F^{mn} (\sigma_{mn})_\alpha^{\phantom{\alpha}\gamma} \delta_a^c+ \frac{1}{2}[\phi^i,\phi^j] (\sigma_{ij})_a^{\phantom{\alpha}c}\delta_\alpha^\gamma +iK^{\hat m}(\sigma_{\hat m 0})_\alpha^{\phantom{\alpha}\gamma}\delta_a^c \right)\theta_{c\gamma}\theta_{b\beta}\displaybreak[0]\\
&\quad+i \epsilon_{\dot a\dot b}\epsilon_{\dot\alpha\dot \beta} \left( \frac{1}{2}F^{mn} (\bar\sigma_{mn})^{\dot\alpha}_{\phantom{\alpha}\dot\gamma} \delta_{\dot c}^{\dot a} + \frac{1}{2}[\phi^i,\phi^j] (\bar\sigma_{ij})^{\dot a}_{\phantom{a}\dot c} \delta_{\dot\gamma}^{\dot\alpha}  +iK^{\hat m} (\bar\sigma_{\hat m 0})^{\dot\alpha}_{\phantom{\alpha}\dot\gamma} \delta_{\dot c}^{\dot a}  \right) \tilde\theta^{\dot c\dot\gamma} \tilde\theta^{\dot b\dot\beta}\displaybreak[0]\\
&\quad +  2i \epsilon_{\dot c\dot b}\epsilon_{\dot\zeta\dot \beta} \left(D^m \phi_{i} - i\delta^m_0 \left(K_i + \frac{1}{\ell} \phi_i\right)\right)(\bar\sigma^{i})^{\dot c c} (\bar\sigma_{m})^{\dot\zeta \zeta} \theta_{c\zeta}\tilde\theta^{\dot b\dot\beta}\displaybreak[0]\\
&\quad-i \epsilon_{\dot c\dot b}\epsilon_{\dot\zeta\dot \beta} \bigg[ \delta_{\dot d}^{\dot c}\Big( \big(- (\bar\sigma^{m})^{\dot\gamma \zeta} D_m\Psi_{+a}^{\dot \alpha} + \frac{2i}{\ell} \Psi_{+a}^{\dot \alpha} (\bar\sigma_0)^{\dot\gamma\zeta}\big)\epsilon^{ac}(\epsilon_{\dot\alpha\dot\eta}\delta_{\dot\gamma}^{\dot\zeta} + \epsilon_{\dot\gamma\dot\eta}\delta_{\dot\alpha}^{\dot\zeta})\\
&\qquad\qquad\qquad\qquad +\big( D_m \Psi_{+a}^{\dot \alpha} (\sigma_m)_{\alpha\dot\alpha}\epsilon^{ac}-i [Z,\Psi_{-a\alpha}]\epsilon^{ac} -[\phi_j, \Psi_{+\alpha}^{\dot a}]  \epsilon_{\dot a\dot e}(\bar\sigma_{j})^{\dot e c}\big)(\bar\sigma^{0})^{\dot\alpha \zeta}(\bar\sigma_0)^{\dot\delta\alpha}(\epsilon_{\dot\alpha\dot\eta}\delta_{\dot\delta}^{\dot\zeta} + \epsilon_{\dot\delta\dot\eta}\delta_{\dot\alpha}^{\dot\zeta}) \Big)\\
&\quad\qquad\qquad\quad+ [\Psi_{+\alpha}^{\dot a}\epsilon^{\alpha\zeta}\delta^{\dot\zeta}_{\dot\eta},(\bar\sigma^{i})^{\dot e c} \phi_i](\epsilon_{\dot a\dot d}\delta_{\dot e}^{\dot c} + \epsilon_{\dot e\dot d}\delta_{\dot a}^{\dot c} ) \bigg] \tilde\theta^{\dot d\dot\eta}\theta_{c\zeta}\tilde\theta^{\dot b\dot\beta}\displaybreak[0]\\
&\quad-i \epsilon_{\dot c\dot b}\epsilon_{\dot\zeta\dot \beta} \bigg[\epsilon^{dc}\Big( \big( D_m\Psi_{+\alpha}^{\dot c} (\bar\sigma^{m})^{\dot\zeta \theta} -\frac{2i}{\ell} \Psi_{+\alpha}^{\dot c}(\bar \sigma_0)^{\dot\zeta\theta} \big)( \epsilon^{\alpha\zeta}\delta^\eta_{\theta} + \epsilon^{\alpha\eta} \delta^\zeta_{\theta})\\
 &\qquad\qquad\qquad\qquad+\big( -D_m \Psi_{+\delta}^{\dot c} (\bar\sigma_m)^{\dot\alpha\delta}  +i[ Z,\Psi_{-}^{\dot c\dot\alpha}]  +  [(\bar\sigma_{j})^{\dot c a}\phi_j,\Psi_{+a}^{\dot\alpha}] \big)(\bar\sigma^{0})^{\dot\zeta \theta}(\sigma_0)_{\alpha\dot\alpha}( \epsilon^{\alpha\zeta}\delta^\eta_{\theta} + \epsilon^{\alpha\eta} \delta^\zeta_{\theta}) \Big)\\
 & \quad\qquad\qquad\quad +[-\Psi_{+a}^{\dot\zeta}\epsilon^{\zeta\eta},(\bar\sigma^{i})^{\dot c e} \phi_i](\epsilon^{ad}\delta_e^c + \epsilon^{ac}\delta_e^d )\bigg] \theta_{d\eta}\theta_{c\zeta}\theta^{\dot b\dot\beta}\displaybreak[0]\\
&\quad -\frac{1}{2} i \epsilon_{\dot c\dot b}\epsilon_{\dot\zeta\dot \beta}\bigg[ \Big(  -iD_mD_p Z  (\bar\sigma^{m})^{\dot\gamma (\zeta} (\bar\sigma^{p})^{\dot\alpha \eta)} \delta_a^e    -i \frac{3}{\ell} \frac{1}{\ell}Z  (\bar\sigma^{0})^{\dot\alpha \eta} (\bar\sigma_0)^{\dot\gamma\zeta} \delta_a^e   \\
&\quad\qquad\qquad\qquad\quad +4\frac{1}{\ell}D_m Z   (\bar\sigma^{m})^{\dot\gamma (\zeta}  (\bar\sigma^{0})^{\dot\alpha \eta)} \delta_a^e   +2 [\Psi_{+d}^{\dot\gamma}\epsilon^{de}\epsilon^{\zeta\eta} ,\Psi_{+a}^{\dot\alpha}]\Big)\delta^{\dot c}_{\dot d}\epsilon^{ac}(\epsilon_{\dot\alpha\dot\eta}\delta_{\dot\gamma}^{\dot\zeta} + \epsilon_{\dot\gamma\dot\eta}\delta_{\dot\alpha}^{\dot\zeta})\\
&\quad\qquad\qquad\qquad+\Big(iD_mD_m Z  \epsilon^{ec}\delta^\eta_\alpha   + [2\Psi_{+d}^{\dot\gamma}\epsilon^{de}\epsilon_{\dot\gamma\dot\alpha}\epsilon^{ac}\delta^\eta_\alpha ,\Psi_{+a}^{\dot\alpha}] -\frac{i}{\ell}\frac{1}{\ell} Z  \epsilon^{ec}\delta^\eta_\alpha \\
&\quad\qquad\qquad\qquad\quad +i [Z,-\frac{1}{2} [\bar Z, Z] \delta^\eta_\alpha ]\epsilon^{ec} -2[ \Psi_{+\gamma}^{\dot b} \epsilon^{\gamma\eta} , \Psi_{+\alpha}^{\dot a}] \epsilon^{ec} \epsilon_{\dot a\dot b} -  [\phi_i, i[\phi_i, Z] \delta^\eta_\alpha  ] \epsilon^{ce}\Big)(\bar\sigma^{0})^{\dot\alpha \zeta}(\bar\sigma_0)^{\dot\delta\alpha}(\epsilon_{\dot\alpha\dot\eta}\delta_{\dot\delta}^{\dot\zeta} + \epsilon_{\dot\delta\dot\eta}\delta_{\dot\alpha}^{\dot\zeta})\delta^{\dot c}_{\dot d}\\
&\quad\qquad\qquad\qquad+ [Z,K^{\hat m} ]\epsilon^{ec}\delta^{\dot c}_{\dot d}(\bar\sigma^{\hat m})^{\dot\delta (\eta}(\bar\sigma_0)^{\dot\alpha\zeta)}(\epsilon_{\dot\alpha\dot\eta}\delta_{\dot\delta}^{\dot\zeta} + \epsilon_{\dot\delta\dot\eta}\delta_{\dot\alpha}^{\dot\zeta})\\
&\quad\qquad\qquad\qquad+\Big( [i[\phi_j, Z]  \epsilon^{\eta\zeta}\delta^{\dot\zeta}_{\dot\eta}, \phi_i](\bar\sigma^{j})^{\dot a (e} (\bar\sigma^{i})^{\dot e c)}+ 2 [\Psi_{+\alpha}^{\dot a}\epsilon^{\alpha\zeta}\delta^{\dot\zeta}_{\dot\eta}, \Psi_{+\beta}^{\dot e}\epsilon^{ec} \epsilon^{\beta\eta} ] \Big)(\epsilon_{\dot a\dot d}\delta_{\dot e}^{\dot c} + \epsilon_{\dot e\dot d}\delta_{\dot a}^{\dot c} )\bigg] \theta_{e\eta}\tilde\theta^{\dot d\dot\eta}\theta_{c\zeta}\tilde\theta^{\dot b\dot\beta}\;.
\end{align*}
\end{scriptsize}%


\section{Details on the perturbative Hamiltonian} \label{pertham}
In this appendix, we provide more details on the perturbative computation of the generators of the $\mathfrak{su}(2|2)\ltimes \mathbb{R}$ algebra. 

With the exception of the tree level case, generally we need to impose the algebraic constraints up to order $k+1$ in order to fix the Hamiltonian at order $k$ and the supersymmetry generators at order $k-2$. The relevant commutation relations for our study are
\beq
\begin{aligned}
&[\mathcal{H-\mathcal{J}},\mathcal{Q}^{a \beta}]= [\mathcal{H-\mathcal{J}},\mathcal{Q}^{\dagger}_{b \alpha}] = 0 \\
&\{\mathcal{Q}^{\dagger}_{a\alpha} , \mathcal{Q}^{b \beta}\} =  \delta_{a}^{b}\, \mathcal{L}_{\alpha}{}^{\beta} + \delta_{\alpha}^{\beta}\, \mathcal{R}_{a}{}^{b} +\frac{1}{2} \delta^{b}_{a}\, \delta^{\beta}_{\alpha} \left( \mathcal{H}-\mathcal{J} \right) 
\end{aligned}
\eeq
since the Lorentz and R-symmetry generators do not get corrected at loop level. After imposing the algebraic constraints, we also need to identify among the remaining unfixed parameters which ones are physical and affect the energies and the ones which are associated to similarity transformations. We can estimate the number of parameters of the similarity transformation. These transformations are generated by an operator $\mathcal{T}(\kappa) $ as
\beq
\mathcal{G}(\kappa) \rightarrow \exp \left( \mathcal{T}(\kappa) \right) \, \mathcal{G}(\kappa)\, \exp(-\mathcal{T}(\kappa)) \,.
\eeq
This operator  $\mathcal{T}(\kappa) $ admits an expansion in the coupling constant $\mathcal{T}(\kappa)  = \sum_{k=1}^{\infty} \kappa^{k} \mathcal{T}_k$ and at each order, $\mathcal{T}_{k}$ must be invariant under Lorentz and R-symmetry, and should preserve the classical energy. Therefore, $\mathcal{T}_{k}$ must be built out of the same tensor structures as the Hamiltonian itself. In particular, $\mathcal{T}_1=0$. This means that at order $\kappa^2$ the Hamiltonian cannot be modified by a similarity transformation (since $[\mathcal{H}_{0},\mathcal{T}_{k}]=0$) but at order $\kappa^4$, it can be changed as
\beq
\mathcal{H}_{4} \rightarrow \mathcal{H}_{4}+[\mathcal{H}_{2},\mathcal{T}_2]\,.
\eeq
The supersymmetry generators are also transformed and at lowest nontrivial order they transform acoording to
\beq
\mathcal{Q}_{2} \rightarrow \mathcal{Q}_{2}+[\mathcal{Q}_{0},\mathcal{T}_2]\,,\quad\quad
\mathcal{Q}^{\dagger}_{2} \rightarrow \mathcal{Q}^{\dagger}_{2}+[\mathcal{Q}^{\dagger}_{0},\mathcal{T}_2]\,.
\eeq
The operator $\mathcal{T}_2$ contains $11$ structures, as in (\ref{ham2}), and we find 7 solutions to the equations $[\mathcal{H}_{2},\mathcal{T}_2]=[\mathcal{Q}_{0},\mathcal{T}_2]=[\mathcal{Q}^{\dagger}_{0},\mathcal{T}_2]=0$. This means that we have a total of four unphysical parameters that can be removed by similarity transformations. Obviously, $\mathcal{H}_2$ itself generates a nontrivial similarity transformation on the supersymmetry generators but it leaves $\mathcal{H}_{4}$ invariant. Therefore, we expect to be able to eliminate three parameters of the Hamiltonian.
In addition, we impose that the action is trivial on cyclic states (this allows to kill three parameters) and use $\mathfrak{su}(2)$ Jacobi identities 
\beq
\PTerm{a b c}{[abc]} = \PTerm{\alpha \beta \gamma}{[\alpha \beta \gamma]} = 0\,, 
\eeq
which allows to further eliminate two parameters.

\pdfbookmark[1]{\refname}{references}
\bibliographystyle{nb}
\bibliography{references}

\end{document}